\documentclass[draft]{agujournal2019}
\usepackage{url} 
\usepackage{lineno}
\usepackage{amsmath}
\usepackage{amsfonts}
\usepackage{booktabs}
\usepackage{siunitx}
\usepackage{multirow}
\usepackage{makecell}

\usepackage[inline]{trackchanges} 
\usepackage{soul}
\usepackage{pgfplotstable}
\usepackage{tikz}
\usepackage{float}
\usepackage{mathtools}
\usepackage{bbm}
\usepackage[ruled]{algorithm2e}
\usepackage{algpseudocode}

\makeatletter
\newcommand{\mathleft}{\@fleqntrue\@mathmargin0pt}
\usetikzlibrary{positioning}
\pgfplotsset{compat=1.7}
\RestyleAlgo{ruled}

\newcommand{\indep}{\perp \!\!\! \perp}
\addeditor{yuchen}
\draftfalse

\journalname{JGR: Solid Earth}

\begin{document}

%
%


\title{Time-Varying Confounding Bias in Observational Geoscience with Application to Induced Seismicity}

%
%




\authors{Yuchen Xiao\affil{1}, Corwin Zigler\affil{1}, Peter H. Hennings\affil{2,3}, Alexandros Savvaidis\affil{2,3}}

\affiliation{1}{Department of Statistics and Data Sciences, University of Texas at Austin, Austin, Texas, USA}
 \affiliation{2}{Jackson School of Geosciences, University of Texas at Austin, Austin, Texas, USA}
 \affiliation{3}{Bureau of Economic Geology, University of Texas at Austin, Austin, Texas, USA}





\correspondingauthor{Yuchen Xiao}{xiao.jack@utexas.edu}



\begin{keypoints}
\item Evaluating causal effects in observational geoscience problems must confront confounding
\item Time-varying confounding is a realistic concern in longitudinal analysis of wastewater disposal on induced seismicity
\item Simulation studies demonstrate the threat of time-varying confounding and alternative estimators are provided to recover unbiased estimates 
\end{keypoints}

%
%

%
%


\begin{abstract} Evidence derived primarily from physical models has identified saltwater disposal as the dominant causal factor that contributes to induced seismicity. To complement physical models, statistical/machine learning (ML) models are designed to measure associations from observational data, either with parametric regression models or more flexible ML models. However, it is often difficult to interpret the statistical significance of a parameter or the predicative power of a model as evidence of causation. We adapt a causal inference framework with the potential outcomes perspective to explicitly define what we meant by causal effect and declare necessary identification conditions to recover unbiased causal effect estimates. In particular, we illustrate the threat of time-varying confounding in observational longitudinal geoscience data through simulations and adapt established statistical methods for longitudinal analysis from the causal interference literature to estimate the effect of wastewater disposal on earthquakes in the Fort-Worth Basin of North Central Texas from 2013 to 2016. 

\end{abstract}

\section*{Plain Language Summary} 

Causal inference, a sub-area of statistics, has gained popularity across other quantitative fields of medicine, epidemiology, and social sciences to support causal interepretation of the analysis of observational studies, but has not been previously explored in geoscience until very recently. We apply a causal framework with the potential outcomes perspective to analyze the effect of the saltwater disposals on earthquakes over a period of time. We specifically focus on unraveling the time-varying confounding bias that naturally occurs in observational geocience problems involving longitudinal data. We provide established methods for the treatment of time varying confounding that has been absent in the induced seismicity literature.

\section{Introduction}
\subsection{Background}


Saltwater disposals (SWDs) have been linked to the recent increase of earthquakes in various regions of the United States \cite{Ellsworth2013InjectionInducedEarthquakes, Frohlich2016ATexas, Grigoratos2020EarthquakesHindcasting, Hennings2019Injection-inducedTexas, Justinic2013Analysis2010, keranen2013potentiallyInducedEarthquakesinOK, Langenbruch2017ResponseRates, mcclure2017identifying, Walsh2015OklahomasDisposal, Weingarten2015High-rateSeismicity}. In Texas, the development of shale hosted hydrocarbon resources in the Permian Basin, Eagle Ford Basin and Barnett Basin has resulted in a rapid expansion in both the number of SWDs and the cumulative injection volume, along with an abrupt increase in the number of earthquakes in respective basins \cite{Hennings2019Injection-inducedTexas, Hornbach2015CausalTexas, Ogwari&Deshon2018DFWEarthquakeSequence, Quinones2019TrackingCatalog, Scales2017ATexas,Zhai2018FluidTexas, gao2019hydrologicalmodelinginDFW}. Of particular importance is the Fort-Worth Basin which hosts Barnett Shale in the North Texas that include most of the Dallas-Fort Worth (DFW) metropolitan area. Although the rate of earthquake activity in the DFW region has decreased since its peak in 2015, the potential linkages to oil and gas activity continue to be a concern and put the social license of developing oil and gas resources in Texas at stake.

\begin{figure}[hbt!]
    \centering
    \includegraphics[width=11cm]{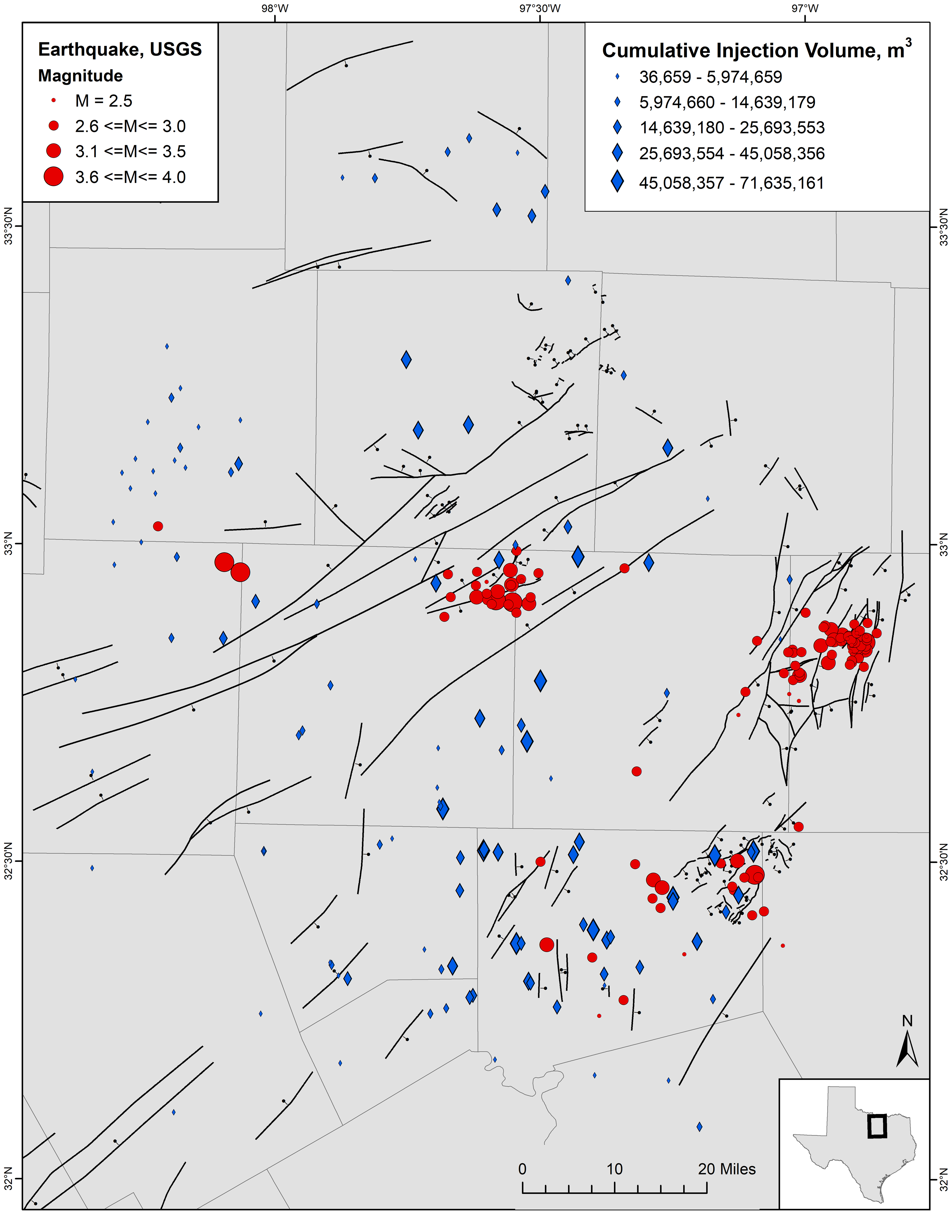}
    \caption{Map shows the study area of Fort-Worth Basin in North Texas. The black lines denote the traces of basement faults from \citeA{Hennings2019Injection-inducedTexas}. The red circles denote the earthquakes from the North Texas Earthquake Study (NTXES) catalog (2008-2018) where the circle size indicates the magnitude of the earthquake.  The earthquake catalog is declustered with Reasenberg's algorithm assuming a magnitude of completeness of 2.5. The blue diamonds denote the SWDs where the size indicates the cumulative injection volume from 2000 through 2017 reported from operators.}
    \label{fig:DFW map}
\end{figure}

In response to this concern, the TexNet Seismological Observatory and the Center for Integrated Seismicity Research (CISR) at The University of Texas at Austin were established to monitor potentially induced seismicity and to better understand the earthquake activities across the State of Texas \cite{Hennings2019Injection-inducedTexas, savvaidis2019texnet}. One of the overarching goals of TexNet-CISR is to improve causative understanding of the relationship between SWDs and onset earthquakes and the quantification of any identified causal relationship. 

The evidence linking induced seismicity to wastewater injection is primarily generated from two domains. First, more refined understanding of the underlying physical processes has been gleaned from numerous physical models that identify significant pore pressure increase, from large-scale SWD activities, reduces the frictional resistance of critically stressed faults, subsequently inducing fault slips \cite{Fan2019BasementEffects, Zhai2018FluidTexas, Keranen2014SharpIncreaseinOKSeismicityRatesBYSWD, Snee&Zoback2016StateofStressinTX}. Second, the obvious temporal and spatial associations between wastewater injection and induced seismicity, combined with detailed statistical analyses of such associations, have corroborated conclusions that SWD is among the causal factors of the observed seismicity.
\cite{Hornbach2016EllenburgerTexas, Fan2019BasementEffects, Langenbruch&Zoback2016HowWillSeismicityinOKRespondtoDecreasedSWD,mcclure2017identifying, savvaidis2020induced}. These sources of evidence should be viewed as complementary; physical models distill enormous amounts of information on the physical processes, but cannot easily provide measurable relationships between specific patterns of SWD and induced seismicity to inform policy changes. Statistical modeling can quantify (with uncertainty) apparent relationships between SWD and induced seismicity, but due to their nonexperimental nature can struggle to parse causal relationships from spurious or coincidental associations. The present paper seeks to improve this latter feature of statistical analysis of induced seismicity related to wastewater injection, focusing, in particular, on statistical methods for causal inference.

Causal inference, a sub-field of statistics, has gained popularity across other quantitative fields of medicine, epidemiology, and social sciences to provide evidence of causality but has not been similarly foregrounded in geosciences \cite{ dominici&Zigler2017GaugingEvidenceofCausalityinAirPollution,  Reich2020Areviewofspatialcausalinferencemethodsforenvironmentalandepidemiologicalapplications, imbens_rubin_2015, perez2018causalgeoscience, massmann2021causalearthscience}. A distinguishing feature of causal inference analysis is that it augments more typical statistical analysis with explicit definition of an inferrential target called a ``causal effect,'' formally defined based on the notion of potential outcomes.  Specifically, a causal effect, which can be defined irrespective of any statistical model, is a comparison between what would potentially happen for a unit (e.g., a particular location in the study region) under different treatments conditions (e.g., injection histories) at a given time \cite{imbens_rubin_2015}. One contribution of the present work is to formalize causal effects that may be of interest in the study of induced seismicity. The virtue of formalizing causal effects in this way is that it supports rigorous charaterization of the identification conditions and assumptions required to reliably estimate such causal effects from available data.  These assumptions are made in service of overcoming  the ``fundamental problem of causal inference'' that we can only observe onsite seismicity under the past observed history of SWD operation \cite{Holland1986statisticsandcausalinference}. 

Previous studies have employed parametric regression models to estimate correlations between annual injection volume and annual seismicity rate in time, interpreting the statistical significance of such correlations as evidence of causation \cite{mcclure2017identifying}. The obvious caveat is that ``correlation does not imply causation''. In particular, if the relationship between a statistical parameter and a precise definition of causal effect is left vague, then even the most sophisticated estimation procedures could provide misguided evidence for causality \cite{Hornbach2015CausalTexas,Fasola2019HFstrategyinfluencetheprobabilityofEarthquakes, mcclure2017identifying, Grigoratos2020EarthquakesForecasting,aldrich1995correlationdoesnotimplycausation,Langenbruch&Zoback2016HowWillSeismicityinOKRespondtoDecreasedSWD, pearl2019seventoolsofcausalinference}. Against the backdrop of potential outcomes, we present similarly-motivated analyses, but with in-depth presentation of a causal inference framework to precisely define causal effects and the attendant assumptions required for estimation.   We offer additional context for evaluating major threats to causal validity for geosciences problems, focusing specifically on the threat of time-varying confounding that can arise when treatment or exposure of interest varies across time (e.g., the monthly injection volume). Aside from a brief discussion of causal assumptions in \citeA{mcclure2017identifying}, existing literature on induced seismicity does not explicitly consider the threat of time-varying confounding. We deploy established methods based on the theory of Marginal Structural Models (MSMs) \cite{robins1997marginal}, a class of methodology designed specifically to adjust for types of time-varying confounding that cannot be accommodated by standard statistical adjustment approaches common to the analysis of time-varying data. A simulation study is designed to emphasize the importance of methods such as MSMs to adjust for time-varying confounding. We apply the method in an empirical analysis of SWD operation and seismicity in the Fort Worth Basin during December 2013 to March 2016 for a 28 month study period.

\section{Methods} \label{methods}
\subsection{Data Assembly and Parameterization} \label{Data Assembly}

Our study area is the Dallas Fort-Worth (DFW) Basin in North-Central Texas. Numerous studies have documented the evolution of earthquake sequences, collected extensive compilations of mapped faults, and conducted numerical simulations of hydrological modeling and fault activation in the area of interest \cite{Hennings2019Injection-inducedTexas, Frohlich2016ATexas, Frohlich2020OnsetArray, Fan2019BasementEffects, Hornbach2016EllenburgerTexas,Scales2017ATexas, Louis&Deshon2018StressOrientationsinDFW, Snee&Zoback2016StateofStressinTX, gao2019hydrologicalmodelinginDFW,Fasola2019HFstrategyinfluencetheprobabilityofEarthquakes, Schoenball&Ellsworth2017Assessmentofspatiotemporalevolutionoffaultactivation}. We refer to above references for complete background information for the study area. We use the North Texas Earthquake Study (NTXES) catalog (2008-2018), collected by the Southern Methodist University (SMU), in this study. We have declustered the earthquake catalog with Reasenberg's algorithm using a magnitude of completeness of 2.5. The SMU catalog had fewer earthquake monitoring stations back in 2008 and the temporary stations have mostly captured the aftershocks, not the main shocks. We use the operator reported SWDs injection volume data in DFW area from 2000 to 2019, available from the Texas Railroad Commission website. Since systematic operation of the network started in 2013 and most earthquakes occurred between 2013 to 2016, we subset both the SWD injection data and earthquake catalog from December 1st of 2013 to March 1st of 2016 for a consecutive 28 months study period. Consequently, a total of 71 earthquakes and 65 SWDs are included in our analysis. The study area is bounded from $32.07$ degree to $33.68$ degree latitude and $-98.38$ degree to $-96.74$ degree longitude. We convert the longitude and latitude coordinates to to Cartesian coordinates with the WGS 84 coordinate reference system. 


Many of the SWDs in the study region are located in close proximity to one another, making it impractical to parse how operation of one SWD may impact nearby seismicity in complete isolation of other co-located SWDs (Section \ref{Conclusion} discusses extensions to address the interdependence among SWD locations). Thus, in service of our focus on causal inference methods for time-varying confounding, we use an agglomerative clustering algorithm with $N = 30$ clusters to partition SWDs into 30 groups and designate each cluster centroid as an observational unit of study. We identify, for each study month, any earthquake that occurred within a distance threshold of 15 kilometers away from each cluster centroid. Earthquakes that are within 15 kilometers of more than one SWD cluster centroid are categorized to the closest cluster to avoid ``double counting'' of earthquakes. We consider a quarterly time scale, and assign each cluster a total quarterly injection volume from all SWDs in the same cluster. We similarly calculate the quarterly count of earthquakes that occurred within 15 kilometers of the cluster centroid. Thus, the analysis data consists of 30 observational units (i.e., 30 SWD clusters) where each cluster has a time series of 7 time points containing the quarterly injection volumes and the total earthquake count during the 28 month period. 

Section \ref{Causaleffecttimefixed} and \ref{IgnorabilityIdentificationTimeFixed} introduce a set of notations and causal identification conditions to define and estimate causal effect under the simplified setting of a time-fixed treatment. Section \ref{MSM} and Section \ref{identificationtimevaryingconfounding} extend the notations and causal identification conditions to accommodate estimation of causal effect in longitudinal analysis with time-varying treatments and time-varying confounding. Section \ref{simulation} presents simulation to illustrate how time-varying confounding might occur in observational geoscience problems and how traditional regression methods with covariate adjustment are insufficient to mitigate time-varying confounding. Section \ref{analysis} describes the longitudinal analysis on the earthquake catalog for the Dallas Fort-Worth region. Section \ref{Conclusion} concludes and discusses future work.

\section{Causal Effects, Identification and Ignorability} 

We offer development for but one of many causal questions one could ask pertaining to induced seismicity: How do differences in injection profile (i.e., the history of quarterly injection volume) lead to (or cause) different patterns of onsite seismicity activity? To formalize this question, let $A_{i}(t)$ denote the injection volume during time $t$ at SWD cluster $i$, with $i=1,2,\ldots, N$ and $t=1,2, \ldots,K$. Denote the entire trajectory of the SWD operation for SWD cluster $i$ up to and including time $t$ with $\bar{A}_i(t) = \{A_i(1), A_i(2),....,A_i(t)\}$. Let $Y_i$ denote the eventual outcome of interest measured at the end of the study period, for example, the cumulative number of earthquakes observed in SWD cluster $i$ at time $K+1$. Denote features of the $i^{th}$ SWD cluster that can vary over time with $\mathbf{L}_i(t)$, and the corresponding time history of such features up to time $t$ with $\bar{\mathbf{L}}_i (t)$. Time-fixed features of each SWD cluster are denoted with $\mathbf{X}_i$.

In a general sense that we will formalize in the subsequent, the goal is to estimate the causal effect of $\bar{A}_i(K)$, the entire history of SWD operation, on $Y_i$, which may require some adjustment for cluster-specific features in $\bar{\mathbf{L}}_i(K)$ and $\mathbf{X}_i$. In the following, we formalize the above causal question using the potential-outcomes framework to (a) clarify salient threats to causal validity of an empirical analysis of SWD and induced seismicity and (b) motivate modern strategies for longitudinal analysis with time-varying confounding. Throughout the paper, we use capital letters to denote random variables (e.g., $A_i$), and lower-case letters to denote a realization of the random variable (e.g., $a_i$). We use bold capital and lower-case letters to denote a vector of random variable and a vector of one realization of the random variable, respectively.

\subsection{Defining the Causal Effect for the Simplified, One-Time Treatment} \label{Causaleffecttimefixed}
For exposition, we introduce key concepts in an intentionally simplified version of the motivating problem with a time-fixed treatment. We formalize a causal effect and detail three identification conditions that are necessary to estimate it with observed data. These concepts and conditions will be extended to the problem of time-varying treatments in Section \ref{identificationtimevaryingconfounding}.

Consider for simplicity a binary summary of SWD operation during the entire study period to classify each SWD cluster as having either ``high'' or ``low'' injection volume with:
\begin{equation} \label{binarytreatment}
   A^*_i = \left\{ \begin{array}{rcl}
1 \qquad &\mbox{for} \qquad&\sum^K_{t = 1} A_i(t) \geq 5,000,000 \mbox{ bbl}\\
0 \qquad& \mbox{for} \qquad& \sum^K_{t = 1} A_i(t) < 5,000,000 \mbox{ bbl} \\
\end{array}\right.
\end{equation}
where $A^* \in [0,1]$ represents $[$low, high$]$ cumulative injection volume, respectively, and can be regarded as a fixed point ``treatment''. We aim to estimate whether high cumulative injection volume causes more earthquakes during the study period relative to what would have occurred if there had been low cumulative injection volume during the study period.        

Owing to the fact that $A_i$ could, in principle, have been either high or low for every SWD cluster, we augment the notation of the observed outcome, $Y_i$, to define the {\it potential outcome} that would have occurred if SWD cluster $i$ had either high or low cumulative injection volume. Let $Y_i^a$ denote the potential earthquake outcome (e.g., the potential cumulative number of earthquakes) for SWD cluster $i$ if it had had cumulative injection volume $A^*=a$. Thus, each cluster has exactly two potential outcomes, $Y_i^0, Y_i^1$, with the former observed ($Y_i^0=Y_i$) only if cluster $i$ was actually observed to have low cumulative injection volume, and the latter observed ($Y_i^1=Y_i$) only if cluster $i$ was actually observed to have high injection volume. The key insight of these definitions is that, even if unobserved, the value $Y_i^{1-A^*_i}$ {\it could} have been observed if, contrary to fact, SWD operation had been different in that cluster. This notion of potential outcomes leads directly to an explicit definition of a causal effect; knowing that $Y_i^0=Y_i^1$ for all $i$ would imply that there is no causal effect of $A^*$ on $Y$ in the sense that earthquake activity would have been exactly the same under alternative SWD operations, whereas having $Y_i^0 \ne Y_i^1$ for some $i$ implies the presence of a causal effect.  This leads to focus on a causal estimand representing the average causal effect (or average treatment effect, ATE) across all units in the study: 
\begin{equation}
  ATE  =\frac{\sum_{i:N}(Y^1_i - Y^0_i)}{N}.\label{SATE}
\end{equation}
Note that the causal effect in (\ref{SATE}) is conceptual, and is defined without regard to any particular statistical model. The fundamental problem for estimating (\ref{SATE}) is that both $Y_i^0$ and $Y_i^1$ can never be simultaneously observed at a single SWD cluster; $Y^0_i$ remains unobserved for every SWD cluster observed to have $A^*_i=1$, and $Y^1_i$ unobserved for every SWD cluster observed to have $A^*_i=0$. Thus, causal inference is inherently a missing data problem of using available observed data to predict the unobserved potential outcomes \cite{Holland1986statisticsandcausalinference}. The key to causal inference will be specifying the assumptions, identification conditions, and estimation strategy to learn about (\ref{SATE}) through information contained in the observed data.
 
\subsection{Identification Conditions for One-Time Treatments} \label{IgnorabilityIdentificationTimeFixed} 
Estimation of the ATE in (\ref{SATE}) typically relies on three standard identification assumptions. The first, which is already implicit in the above notation, is the Stable Unit Treatment Value Assumption (SUTVA), principally stating that a SWD cluster's earthquake outcome depends only on its \textit{own} injection volume, and not the injection volume from other SWD clusters \cite{ImbensandRubin1996, imbens_rubin_2015}.  The current development assumes that SWD clusters are sufficiently separated for the SUTVA assumption to hold.  Such an assumption may not hold in studies of induced seismicity, which would require nontrivial extensions to formalize causal effects. We revisit this point in Section \ref{Conclusion}, noting here that analogous assumptions have been made in previous statistical investigations of induced seismicity \cite{mcclure2017identifying, Ellsworth2013InjectionInducedEarthquakes}. Another technical component of SUTVA, sometimes termed {\it consistency}, assumes that outcomes only depend on the value of $A^*$, and not the manner in which the level of injection volume was achieved, implying that $Y_i=Y_i^{a^{obs}}$ whenever $A^*_i = a^{obs}$.

Finally, the positivity condition requires that all SWD clusters are eligible to sustain injection activities, that is, there are no locations with features such that $A_i^*=1$ (or $A_i^*=0$) with probability 1. Taken together, inference for the ATE in (\ref{SATE}) under the three identification conditions  would be trivial in the idealized case where SWD clusters were randomly assigned to either high or low cumulative injection volume. Since randomization would ensure comparable background characteristics among high vs. low injection volume locations, outcomes observed among $\{Y_i; A^*_i=0\}$ would support a reliable estimate of average seismic activity whenever cumulative injection volume is low, and outcomes observed among $\{Y_i; A^*_i=1\}$ would support a reliable estimate of average seismic activity whenever cumulative injection volume is high. Unfortunately, the observational nature of data observed in geosciences is relatively sparse in scale and the decisions about SWD operation are not randomized, meaning that systematic differences between clusters with $A_i^*=0$ and those with $A_i^*=1$ are likely to {\it confound} any comparisons among their observed outcomes. The assumption of {\it ignorable treatment assignment}, also called the {\it no unmeasured confounding} assumption or {\it conditional exchangeability} encodes the assumption that observations of $Y^0_i$ for clusters with $A^*_i=0$ could anchor inferences about the unobserved values of $Y^0_j$ for clusters with $A^*_j=1$ (and {\it vice versa}), provided that any systematic differences among clusters that are related to the earthquake outcomes are conditioned upon (or adjusted for) in the estimation of (\ref{SATE}) \cite{imbens_rubin_2015}:
\begin{equation}\label{ignorability}
    \{Y^0_i, Y^1_i\} \indep A^*_i | \textbf{X}_i 
\end{equation}

The less formal operational definition of (\ref{ignorability}) is that $\textbf{X}_i$ is a confounder that must be conditioned upon to satisfy (\ref{ignorability}) if it is jointly associated with both $A_i^*$ and $Y_i$.  Possible confounders that might be required to satisfy (\ref{ignorability}) in studies of induced seismicity include but are not limited to hydrocarbon deposits, geologic elevation, hydrocarbon producing wells, and geological faults, all of which may be inputs into decisions about SWD operation and also predispose certain areas to more earthquake activity \cite{Hennings2019Injection-inducedTexas, Snee&Zoback2016StateofStressinTX}. If clusters with low injection volume tend to exhibit different predisposition to earthquake activity than clusters with high injection volume, then, for (\ref{ignorability}) to hold, features encoding such predisposition must be encoded within $\mathbf{X}_i$.  Otherwise, the observed $Y^0_i$ in $A^*_i=0$ clusters may not anchor reliable inference about the unobserved values of $Y^0_j$ in $A^*_j=1$, and estimates of the ATE in (\ref{SATE}) will be {\it confounded}. 

Importantly, the ignorability assumption in (\ref{ignorability}) cannot be verified empirically. If we do not believe the ignorability assumption holds, we should not believe that our statistical analysis is estimating a causal effect. This underscores the necessity of understanding, and collecting data on, factors that are thought to relate to SWD operation and cumulative number of earthquakes in order to maximize the belief in the ignorability assumption. Note that specification of the ATE and its identifiability conditions have not relied on any particular statistical model or method for estimation. This is a key feature of potential-outcomes approach to formalize the quantity of interest and key assumptions with potential outcomes, separate from any downstream model that may be used for estimation. This precise mathematical formalization of a causal effect and what exactly would be required to estimate it from observed data is a foundation from which contextual knowledge of the problem can be used to judge the validity of any analysis attempting to characterize causality. 

With a precise definition of causal effect and satisfaction of the identifying assumptions, a common estimator for the ATE relies on inverse probability of treatment weighting (IPTW):
\begin{equation}
    \widehat{ATE} = \frac{1}{N_1} \sum_{A^*_i =1} \frac{I(A^*_i=1)}{f(A^*_i|\mathbf{X}_i)}Y_i - \frac{1}{N_0}\sum_{A^*_i = 0} \frac{I(A^*_i=0)}{(1-f(A^*_i|\mathbf{X}_i))}Y_i \mbox{,} \quad \quad  N_a = |\{i :A^*_i = a \}| \label{iptw_estimator}
\end{equation}
which is unbiased and consistent for the ATE in (\ref{SATE}) \cite{imbens_rubin_2015}. The quantity in the denominator is called the {\it propensity score}, encoding the conditional probability that cluster $i$ with covariate vector $\mathbf{X}_i$ would have high injection volume. When $A^*_i$ is a binary variable, a logistic regression model is commonly applied to estimate the propensity score.

We forego an example of estimating the causal effect of the single time point treatment $A^*$ with (\ref{iptw_estimator}) and defer estimation details to the time-varying case of primary interest in the following sections. The main point of this section is to establish basic concepts and conditions in this simplified case in service of their further developments in the case of time-varying treatments and to note that, whether explicitly stated or not, some version of these identifiability conditions are required in order to interpret comparisons in observational data as causal effects \cite{mcclure2017identifying}.

\subsection{Marginal Structure Models for Time-Varying Treatments} 
\label{MSM}

We now return to the case of primary interest, which is estimating the causal effect of the entire history of SWD operation, $\bar{A}_i(K)$, on $Y_i$, taken here to be the cumulative number of earthquakes observed at cluster $i$ by time $K+1$. As in the one time point case, we define potential seismicity outcomes under different SWD operation, where such operation is now denoted by $\bar{A}(K)$.  Note that we also dispense of the simplification where operation is either ``high'' or ``low'' and regard $A_i(t)$ as a continuous measure of injection volume.  Let $Y_i^{\bar{a}}$ denote the potential outcome for SWD cluster $i$ had it exhibited injection history $\bar{A}_i(K)=\bar{a}$. Then, causal effects for time-varying treatments are defined as comparisons between earthquake outcomes under different SWD operation histories. For example, $E(Y_i^{\bar{a}}) - E(Y_i^{\bar{a}'})$ is the average causal effect if all clusters had followed injection history $\bar{a}$ compared to $\bar{a}'$. With $K$ time points and continuously-scaled SWD injection volumes, this approach immediately confronts the practical impossibility of characterizing effects of every possible SWD operation history. To address this problem, we turn to Marginal Structural Models (MSM) \cite{robins2000marginal, imbens_rubin_2015}.

Rather than attempt to estimate causal effects defined by any possible comparison of SWD operation histories, MSMs offer a way to specify a structural relationship between certain features of injection histories and potential seismicity outcomes. For example, a common MSM, and the one that we will use in this work, specifies that the potential seismicity outcomes, $Y_i^{\bar{a}}$, do not depend on the precise values of the entire vector $\bar{a}$ but rather on the cumulative injection volume  during the study period implied by a particular operation history, denoted $cum(\Bar{a})$. That is, we specify a relationship of the form: 
\begin{equation}
    E(Y^{\bar{a}}) = exp(\beta_0 + \beta_1 cum(\bar{a})) \label{CausalMSM}
\end{equation}
where $cum(\bar{a}) = \sum_{t=1}^K a_t$. The MSM in (\ref{CausalMSM}) encodes a structural dependence between the expected cumulative number of earthquakes under any SWD operation history and the cumulative injection volume implied by that history. For example, if the cumulative number of earthquakes were assumed to follow a Poisson distribution, $exp(\beta_1)$ would denote the relative rate of the cumulative number of earthquakes attributable to a one unit increase in cumulative injection volume. In general, an MSM could be specified to encode structural dependencies for other features of SWD operation history. Our choice of cumulative injection volume is motivated by the induced seismicity literature that typically assumes the cumulative number of earthquakes follows a Poisson-distributed random variable and regress the cumulative number of earthquakes onto the cumulative injection volume \cite{mcclure2017identifying, Grigoratos2020EarthquakesForecasting, Snee&Zoback2016StateofStressinTX}. In the same manner that the ATE in (\ref{SATE}) represents a precise definition of a causal effect in the time fixed case, so does the specification of the MSM in (\ref{CausalMSM}) for the time-varying case. Since the MSM in (\ref{CausalMSM}) is specified on the potential outcomes directly, it cannot be readily estimated with observed data. The key challenge remains to devise a strategy to estimate $\beta_1$.

\subsection{Identification Conditions for Time-Varying Treatments}\label{identificationtimevaryingconfounding}

As in the time-fixed setting, estimation of the causal effect in (\ref{CausalMSM}) relies on several standard identification conditions. The first two are essentially technical extensions of SUTVA and positivity (defined in Section \ref{IgnorabilityIdentificationTimeFixed}) to encode analogous concepts, namely, that cumulative number of earthquakes at SWD cluster $i$ only depends on the SWD operation history at SWD cluster $i$ (and not other SWD clusters) and that there is some positive probability that any SWD operation history of interest actually occurs. We forego the technicalities of these assumptions for ease of exposition and refer interested readers to \citeA{robins2000marginal} and \citeA{imbens_rubin_2015}.

The more salient identification condition when studying the effect of time-varying SWD operation on seismicity relates to the notion of {\it time-varying confounding}. While it may appear that one could simply define the cumulative injection volume for every SWD cluster and then estimate the causal effect of that quantity (which would closely parallel the simplified analysis put forth in Section \ref{IgnorabilityIdentificationTimeFixed}), the time-varying nature of SWD operation introduces the possibility (and, indeed, the likelihood), that SWD injection histories are informed in part by intermediate information that becomes available during the study period. If that intermediate information is a) impacted by past operation decisions; b) informs future operation decisions; c) relates to the outcome of interest (possibly through dependence induced by some unmeasured factors), then a simple regression analysis comparing observed levels of cumulative injection volume will confront what is known as {\it treatment-confounder feedback bias} \cite{longitudinaldataanalysis}.

\begin{figure}[H] 
     \centering 
 \begin{tikzpicture}[
 roundnode/.style={circle, draw=black!, very thick, minimum size=12mm, inner sep=0pt},
 squarednode/.style={rectangle, draw=black!, very thick, minimum size=12mm, inner sep=0pt},
 ]
 \node[squarednode]      (Zt)          {$A(t)$};
 \node[roundnode]        (Yt)         [right=of Zt]   {$L(t)$};
 \node[squarednode]      (Ztplus1)       [right=of Yt]   {$A(t+1)$};
 \node[roundnode]        (Ytplus1)       [right=of Ztplus1] {$L(t+1)$};
 \node[squarednode]       (Ztplus2)       [right=of Ytplus1] {$A(t+2)$};
 \node[roundnode]        (Ytplus2)       [right=of Ztplus2] {$L(t+2)$};
 \node[roundnode]       (Y)       [right=of Ytplus2] {$Y$};
 \draw[ultra thick,green,->] (Zt.east) -- (Yt.west);
 \draw[ultra thick,cyan,->] (Yt.east) -- (Ztplus1.west);
 \draw[ultra thick,green,->] (Ztplus1.east) -- (Ytplus1.west);
 \draw[ultra thick,cyan,->] (Ytplus1.east) -- (Ztplus2.west);
 \draw[ultra thick,green,->] (Ztplus2.east) -- (Ytplus2.west);
 \draw[ultra thick, orange, ->] (Ytplus2.east) --(Y.west);

 \draw[ultra thick,red!60,->] (Zt.north) to [out=20,in=150] (Ztplus1.north);
 \draw[ultra thick,red!60,->] (Ztplus1.north) to [out=20,in=150] (Ztplus2.north);

 \draw[ultra thick,black!30!green,->] (Zt.north) to [out=25,in=150] (Y.north);
 \draw[ultra thick,black!30!green,->] (Ztplus1.north) to [out=25,in=150] (Y.north);
 \draw[ultra thick,black!30!green,->] (Ztplus2.north) to [out=25,in=150] (Y.north);

 \draw[ultra thick,pink,->] (Yt.south) to [out=-35,in=-150] (Ytplus1.south);
 \draw[ultra thick,pink,->] (Ytplus1.south) to [out=-35,in=-150] (Ytplus2.south);

 \draw[ultra thick, orange, ->] (Yt.south) to [out=-35,in=-140] (Y.south);
 \draw[ultra thick, orange, ->] (Ytplus1.south) to [out=-35,in=-140] (Y.south);

 \end{tikzpicture}
 \caption{The DAG shows a complicated case of time-varying confounding with application in induced seismicity. The $A(t)$ denotes the monthly injection volume at time $t$, $L(t)$ denotes an intermediate measure of seismic propensity at time $t$ and $Y$ is the number of cumulative earthquakes at time $K+1$. The arrows represent established causal pathways in induced seismicity.}
 \label{long dag 1}
 \end{figure}
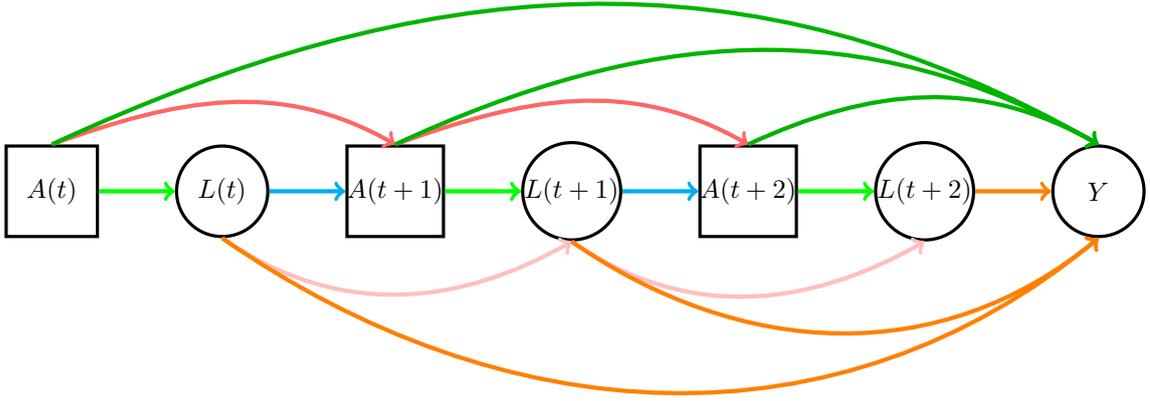

To illustrate, Figure \ref{long dag 1} depicts a directed acyclic graph (DAG) to structural relationships among study quantities. The arrows in the DAG represent presumed causal relationships, where the quantity with the inward edge is caused by the quantity where the arrow originates. Here, $L_i(t)$ represents any information available to the operator that might dictate decisions about operation in a manner that depends on previous operation decisions, while also having some bearing on future operation decisions and cumulative number of earthquakes. We refer to such $L(t)$ generically as intermediate indications of seismic propensity, which may include information on the presence or largest magnitude of earthquakes in a previous month, apparent earthquake activity at nearby locations, or regulatory rules that mandate certain production patterns.

More specifically, the structural assumptions encoded in Figure \ref{long dag 1} specify that $A(t)$ depends on $A(t-1)$ (i.e., red arrows), and $A(t)$ for $t = 1,2,...,K$ can affect $Y$ (i.e., dark green arrows). $L(t)$ is a direct causal consequence of $A(t)$ (i.e., light green arrows) and it {\it also} has bearing on $A(t+1)$ (i.e., blue arrows). This would be the case if, for example, operators reduce injection activity in response to the emergence of intermediate indication of seismic propensity or gradually increase injection activity in the absence of any intermediate indication. In addition, $L(t-1)$ could exert partial influence on $L(t)$, especially when earthquake sequences start to develop (i.e., pink arrows) and $L(t)$ for $t = 1,2,...,K$ are related to $Y$ (i.e., orange arrows), possibly through dependence induced by some unmeasured factors. Omitted from the DAG for simplicity are possible dependencies on time-fixed features, $\mathbf{X}$, which could have outward arrows to any of the depicted quantities. 

By the definition given in Section \ref{IgnorabilityIdentificationTimeFixed}, $L(t)$ is a confounder for the effect of $A(t+1)$ on $Y$. A natural inclination based on the discussion of the simplified time-fixed example might be to then conduct a regression analysis that adjusts for the time-varying confounder, $L(t)$, when estimating the effect of $\bar{A}(t)$ on $Y$. However, such adjustment can actually induce additional bias. While $L(t)$ confounds the relationship between $A(t+1)$ and $Y$, it is also a causal consequence of $A(t)$. This introduces what is called {\it treatment-confounder feedback}. When treatment-confounder feedback is present, conditioning on $\bar{L}(t)$ via regression adjustment would induce a certain type of ``collider bias'', where ``collider'' is used to denote a variable where two inward arrows ``collide,'' as is the case with $L(t)$ in Figure \ref{long dag 1} \cite{cole2010coliderillustrating,elwert2014colliderendogenous}. Regression adjustment on such a collider would misattribute some of the very effect we wish to estimate away from the treatment variable and to the adjustment for the time varying confounder. While we will illustrate this phenomenon in a simulation study below, we refer interested readers to Chapter 20 of \citeA{longitudinaldataanalysis} and \citeA{robins2000marginal,young2010relation,cole2010coliderillustrating,elwert2014colliderendogenous} for a detailed review of DAGs and further explanation of time-varying confounding with treatment-confounder feedback.

\subsection{Sequential Ignorability} \label{sequential ignorability}
The key identification condition to accommodate time-varying confounding with treatment-confounder feedback is the {\it sequential ignorability} assumption. This assumption relies on viewing the injection volume at time $t = 1,2,...,K$ as a sequential decision process; decisions about injection volume at time $t$ are made with all information available at that time about past injection volume and past values of measured time-fixed and time-varying confounders. The assumption states that all seismicity-relevant information available to the operator is encoded by past observed injection volume and confounder values such that, conditional on these values, injection volume at time $t$, $A(t)$, is effectively randomized in the sense that it is independent of potential cumulative number of earthquakes. Sequential ignorability is formally defined as: 

\begin{equation}  \label{sequentiallignorabilityequation}                                
\begin{split}                                      
    Y_i^{\bar{a}} \indep A_i(t) | \bar{A}_i(t-1) = \bar{a}_i(t-1), \bar{L}_i(t-1) = \bar{l}_i(t-1) \\ \mbox{ for all } \bar{a}_i(t-1) \mbox{ and } \bar{l}_i(t-1)              
    \end{split}
\end{equation} 


This assumption extends the ignorability assumption from Section \ref{Causaleffecttimefixed}, essentially casting the sequential SWD operation decision as a sequentially-randomized experiment. Just as in the case of a time-fixed treatment, the sequential ignorability assumption cannot be empirically verified. The main task then is to maximize the chance that the sequential ignorability assumption can plausibly hold by understanding as best as possible what factors dictate operation decisions and influence the eventual earthquake outcome of interest and collect data on as many relevant time-fixed and time-varying confounding variables as possible. Ultimately, the causal validity of results must be judged relative to the plausibility of the sequential ignorability assumption \cite{longitudinaldataanalysis}.

\subsection{Estimation with Inverse Probability of Treatment Weighting} \label{estimateIPTW}

\citeA{robins2000marginal} showed, under the three identifiability conditions, consistent estimators for the causal parameters $\beta_0$ and $\beta_1$ from (\ref{CausalMSM}) can be obtained from observed data with weighted Poisson regression: 
\begin{equation} \label{naivemodel}
      E(Y|\bar{A} = \bar{a}) = exp(\eta_0 + \eta_1 cum(\bar{a}))
\end{equation} where the weights are estimated with IPTW. The time-fixed IPTW defined in Section \ref{IgnorabilityIdentificationTimeFixed} is extended to a product of sequential treatment probabilities across time $t=1,2,\ldots,K$ for IPTW in time-varying treatments:                          

\begin{align}
    SW(K) &= \prod^K_{t=1} \frac{f[A(t)|\bar{A}(t-1)]}{f[A(t)|\bar{A}(t-1), \bar{L}(t-1)]} \label{sw}
\end{align}

The denominator of (\ref{sw}) is the probability density of having observed injection volume, $A(t)$, at time $t$, conditional on observed past histories of $\bar{A}(t-1)$ and $\bar{L}(t-1)$. The numerator is a weight-stabilizing factor specified to alleviate the threat of inflated variance estimates of $\beta_1$ owing to extreme values of the IPTW \cite{longitudinaldataanalysis}. Because the treatment (i.e., the injection volume at time $t$) is now a continuous variable instead of a binary variable as in Section \ref{IgnorabilityIdentificationTimeFixed}, the denominator is more formally referred to as a {\it generalized propensity score} \cite{hirano2004generalizedpropensity}. More details on the weight-stabilizing factor and the use of IPTW to derive asymptotically unbiased estimates in the presence of time-varying confounding can be found in \citeA{longitudinaldataanalysis} and \citeA{robins2000marginal}.

\section{Simulation Study to Illustrate Time-Varying Confounding and MSMs} \label{simulation}

We offer a simulation study that generates longitudinal data to reflect the treatment-confounder feedback situation described above. The analysis goal is to evaluate the average causal effect of cumulative injection volume during the study period on the cumulative number of earthquakes in a way that controls for the interplay between intermediate indications of seismic propensity and operator's decision making. We illustrate how, in a setting such as this, standard regression adjustments will produce biased estimates of causal effects but an MSM can recover unbiased estimates of the causal effect.

\subsection{Data Generation}                  
We generate $N=50$ SWD clusters and set $K=8$ time points to mimic the attributes of the DFW datasets, where $A(t)$, $Y$, and $L(t)$ are defined as above to represent, respectively, continuous measures of injection volume, the cumulative number of earthquakes during the study period, and binary indicators of the presence/absence of intermediate seismic indicators. To generate the dependence between $L(t-1)$ and $L(t)$ that constitutes treatment-confounder feedback, the data generation involves simulation of an unobserved quantity, $U$, which encodes a time-fixed latent seismic risk that influences both $L(t)$ for all $t \in (1,K)$ and $Y$. That is, $U$ is a simulation device that governs such dependence but should be regarded as completely unobserved for analysis. Figure \ref{long dag 2} in the Appendix describes the data generation and the dependence on the unobserved $U$, which corresponds to the DAG in Figure \ref{long dag 1} if $U$ were omitted but its induced dependencies among $L(t)$ and $Y$ remained. In practice, the analysis must control for the underlying relationships through careful adjustment for the observed quantities $L(t)$ and $A(t)$. For simplicity in the simulation, $U$ takes on integer values in $1, 2, ... ,10$ with higher values encoding higher latent seismic risk. 

The procedure for generating $M=2000$ simulated datasets appears in Algorithm \ref{algorithm1} in \ref{datagenerationalgorithm}. In particular, we specify one unit increase in $U$ and $A(t)>1000$ correspond 1.15 times higher risk (i.e., exp(0.14)) and 3 times higher risk (i.e., exp(1.1)) of an intermediate seismic indicator in (\ref{bernP}), respectively. We assume that, in absence of such a seismic indicator, SWD operation would increase by an average of 15 bbl over the previous month, while the presence of intermediate seismic indicators would reduce operation by 40 bbl on average. Lastly, we define the relative rate of number of earthquakes caused by one thousand bbl increase cumulative injection volume (i.e., $\sum_{t=1}^K \textbf{A}_t$) to be 1.359 (i.e, exp(0.5)) in (\ref{Ysimulated}) such that each additional thousand bbl of cumulative injection volume increases the risk of earthquakes by 36\%.

\subsection{Estimating the Causal Effect: Different Strategies for Adjustment} \label{diffentadjustment}

We implement three Poisson regressions to estimate the effect of cumulative injection volume on the number of cumulative earthquakes. The first is a naive Poisson regression, (\ref{naivemodel}), that regresses the number of cumulative earthquakes on the cumulative injection volume without any adjustment. The naive Poisson regression makes no attempt to adjust for intermediate indications of seismic propensity which is a time-varying confounder, therefore the estimated parameters are susceptible to confounding bias. 

Next, we fit a multivariate Poisson regression in (\ref{adjustconfoundermodel}) that adjusts for the observed intermediate indicators of seismic propensity. Although (\ref{adjustconfoundermodel}) adjusts for the observed intermediate confounder, it does not account for the sequential decision making process which creates treatment-confounder feedback which might induce ``collider bias" in time-varying treatments analysis.

\begin{equation}
    E(Y|\bar{A} = \bar{a}) = exp(\gamma_0 + \gamma_1 cum(\bar{a}) + \gamma_2 cum(\bar{l})) \label{adjustconfoundermodel}
\end{equation}

Finally, we use the MSM in (\ref{CausalMSM}), estimated with a weighted Poisson regression with IPTWs as described in Section \ref{estimateIPTW}. Specifically, we use the expression for $SW(K)$ in (\ref{sw}), where $f(\cdot)$ in both the numerator and denominator are linear regressions, and these models are simplified to depend only on the immediately preceding time point $t-1$ and not the entire history. While more complex statistical/ML models could be implemented to model the possible non-linear and complex dependencies \cite{zhang2023propensity}, these simpler models are employed for simplicity and to correspond directly to the data generating mechanism of this simulation study. Variance estimates from models (\ref{naivemodel}) and (\ref{adjustconfoundermodel}) are the maximum likelihood estimates, and those from the weighted Poisson regression employ the robust sandwich variance estimator for conservative variance estimation when IPTW are involved \cite{freedman2006sandwich, wooldridge2007iptwVariance,white1980heteroskedasticity, cribari2011vcovHC4}.



\subsection{Simulation Results}

\begin{figure}[H] 
    \centering
    \includegraphics[width=13cm]{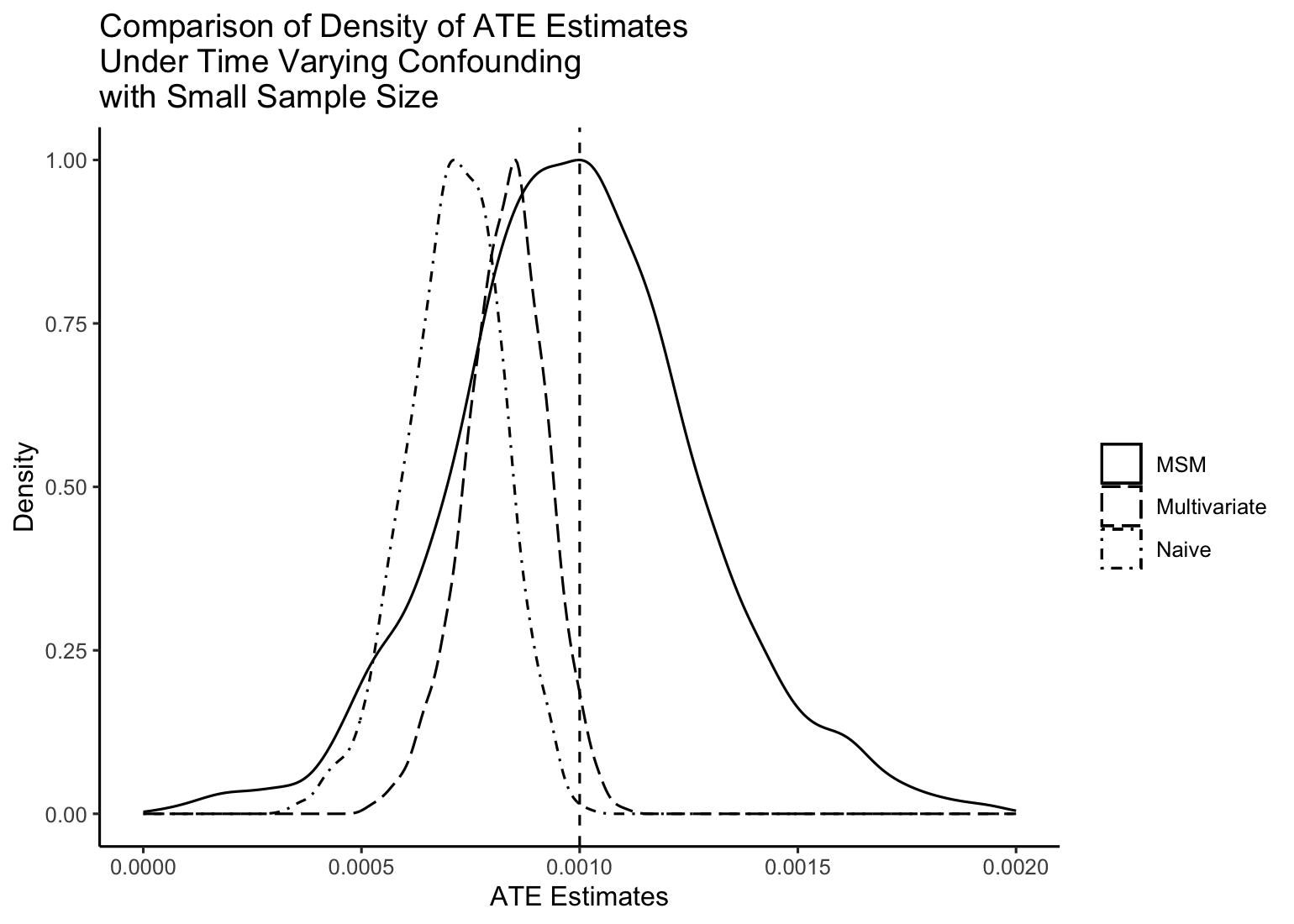}
    \caption{The figure shows the comparison of 2000 simulations of ATE estimates between three Poisson regressions. The solid line is the density for the MSM with \textbf{ipw}, the dot dash line is the density for the naive Poisson regression, and the long dash line is the density for the multivariate Poisson regression. The dashed vertical line displays the true causal parameter used in the data generation. The MSM with \textbf{ipw} recovers the unbiased estimate of the true causal parameter whereas the other two models produce estimates with pronounced bias.}
    \label{fig:Simulationestimatecomparison}
\end{figure}

For each generated dataset, we implement the three approaches described above and collect the estimated coefficients, $\beta_1, \eta_1 , \gamma_1$, respectively. Figure \ref{fig:Simulationestimatecomparison} displays the comparison of the estimated regression coefficients. Each density represents the distribution of estimated regression coefficients across the 2000 generated datasets. The dashed vertical line marks the true causal parameter used to generate the simulated data.

The naive Poisson regression, (\ref{naivemodel}), makes no attempt to adjust for time-varying confounding, and systematically underestimates the true causal parameter. The multivariate Poisson regression, (\ref{adjustconfoundermodel}), adjusts for confounding with a summary of the observed values of the time-dependent confounder. However, adjusting the summary of observed time-varying information does not take into account the sequential decision making process and misattributes some of the true causal effect of $A(t)$ to the adjustment for $cum(\bar{l})$, therefore, the multivariate Poisson regression also systematically underestimates the effect. In contrast, the MSM, (\ref{CausalMSM}), estimated with IPTW recovers the unbiased estimates of the true causal effect parameter (i.e., the mode of the estimates across 2000 generated datasets coincides with the dashed vertical line), albeit with more variability in point estimates. Note that, in general, the discrepancy between the various model estimates would depend on the strength of associations that dictate the treatment-confounder feedback and the magnitude of the true causal effect. 

Table \ref{simulationtabl} provides numerical results for the average bias, standard error estimates, and empirical coverage of 95\% intervals across the three methods. Notice in particular that confidence intervals from the the naive and multiple Poisson regression models cover teh true value far less than 95\% of the time, while the MSM with IPTW has coverage much more closer to the nominal 95\%.  Furthermore, the larger standard error for MSM is associated with the use of sandwich variance estimator for conservative variance estimation.

\begin{table}[h!]
\centering
\begin{tabular}{m{5cm} m{2cm} m{2cm} m{2cm}} 
\hline
Models &  Average Point Estimates & Average Standard Error & Coverage\\
\hline
   Naive Regression Model  & 7.1e-04 & 8.65e-05 & 0.22 \\
   Multiple Regression Model  & 8.3e-04 &  6.25e-05  & 0.36 \\
   MSM with \textbf{ipw}  & 9.9e-04  & 2.24e-04 & 0.91 \\
\hline
\end{tabular} 
\caption{The average point estimate of regression coefficients, in thousands bbl, across 2000 datasets is shown, along with the average of standard errors for each Poisson regression, respectively. Furthermore, the biased estimation in point estimate has resulted in low coverage rate for naive and multiple Poisson regressions.}
\label{simulationtabl}
\end{table}

\section{Analysis of Induced Seismicity in the DFW During 2013-2016} \label{analysis}

Recall from Section \ref{Data Assembly}, $A_i(t)$ denotes quarterly injection volume and $Y_i$ represents the cumulative number of earthquakes from Dec 1 2013 to March 1 2016. We take the intermediate indicator of seismic propensity, $L_i(t)$, to be whether or not any earthquake occurred at cluster $i$ during quarter $t$, under the premise the presence/absence of earthquakes during the study period can impact SWD operation. We deploy the Poisson regressions put forth in Section \ref{diffentadjustment} to estimate the relative risk on cumulative number of earthquakes for every unit increase in cumulative injection volume. Table \ref{resultstable} displays the estimated relative rates and standard errors, scaled to represent the relative rate of earthquakes associated with an increase in 1 MMbbl in cumulative injection volume over the study period. 

\begin{table}[H]
\centering
\begin{tabular}{m{2cm} m{2cm} m{2cm} m{2cm} m{2cm} m{2cm}}
\hline
Models & Parameter & Estimated Relative Risk in 1 MMbbl &  Standard Error & Wald Test Statistic &  P-values \\
\hline
   Naive    & $\gamma_1$ & 1.0056 & 1.089e-02 & 0.521    & 0.6026  \\
   Multivariate  & $\eta_1$ & 1.0307 & 1.201e-02 &  2.518 & 0.0118   \\
   MSM   & $\beta_1$ & 1.0278 & 3.344e-02 & 2.896   &  0.0037 \\
 \hline
 \end{tabular}
 \caption{The results table shows the estimated relative risk in 1 MMbbl, $exp(\beta_1 * 1e06)$, instead of the estimated regression coefficients, $\beta_1$, for the ease of interpretation. }
 \label{resultstable}
 \end{table} 

We estimate the relative risk to be 1.0056, 1.0307 and 1.0278 on the cumulative number of earthquakes for every additional 1 MMbbl increase in cumulative injection volume for naive Poisson regression, multivariate Poisson regression and MSM, respectively. This corresponds to an estimated 3$\%$ increase in risk for 1 MMbbl increase in cumulative injection volume. We note the discrepancy between the naive estimate and the MSM estimate indicates potential threats of time-varying confounding. We use Wald test statistics (i.e., $Z = \frac{\hat{\beta}}{SE\{ \hat{\beta}\}}$) to calculate p-values associated with the estimated coefficients based on estimated standard error. Table \ref{resultstable} shows we found a statistically significant relationship between cumulative number of earthquakes and cumulative injection volume. We conduct a sensitivity analysis on the number of clusters and summarise the results in \ref{realdatasensitivity}.

\section{Discussion, Conclusions, and Future Work} \label{Conclusion}

\subsection{Conclusion}

The quantification of the temporal relationships between injection activity and induced seismicity using observational geosciences data is likely to confront the threat of time-varying confounding.  To illustrate this and offer a strategy to protect against this threat to validity, we formulate an investigation of induced seismicity using the potential outcomes framework to explicitly define the casual estimand and provide required identification conditions. Importantly, we discuss the critical role of the ignorability assumption in mitigating confounding bias for both time-fixed treatment and time-varying treatments. Simulation studies illustrate that traditional regression models with standard covariate adjustment for time-varying features are unable to recover unbiased estimates of the casual estimand in the presence of time-varying confounding, while MSMs estimated with IPTW can recover accurate estimates of causal effects. 

Our analysis on DFW using the SMU catalog shows the results obtained from naive Poisson regression and multivariate Poisson regression are different from that of MSM with IPTW, providing evidence for the presence of time-varying confounding. The results obtained from MSM with IPTW concludes significant statistical risk of SWDs on induced seismicity, with uncertainty quantification, in the DFW region between December 2013 to March 2016, consistent with the results from prior studies. The estimated relative risk is 1.0278, which corresponds to an estimated 2.78$\%$ increase in risk for each additional 1 MMbbl increase in cumulative injection volume. Note the estimated relative risk depends on the choices on the relatively coarse temporal scales, the magnitude of completeness, and the number of clusters implied by the agglomerative clustering algorithm. 


We emphasize that causal validity from empirical analysis on observational data crucially relies on identification conditions. For instance, in hydrological modeling, the validity of the modeling results depends on whether to assume the subsurface is homogeneous or heterogeneous for porosity and permeability. Whether one uses finite difference or finite element method has less impact on validity of the modeling results if one assumes the subsurface is homogeneous. Analogously, using more complex statistical/ML models to estimate the effect of $\bar{A}$ on $Y$ does not correct the time-varying confounding bias that stems from inadequate adjustment of time-dependent confounders. We purposely define the causal estimand and establish causal identification conditions before implementing any estimation methods which enables obtaining an explicit quantity for the association between induced seismicity and SWDs. We argue a more general causal formulation of the problem with the potential outcomes perspective could faithfully reflect the complexity of the problem and improve the clarity and transparency regarding the most important tenets for discerning whether empirical statistical analyses provide evidence of causality between SWD and seismicity \cite{imbens_rubin_2015, carone&Dominici2020TheRoleofCausallyDrivenDataScience}.

\subsection{Methodological Comparison with Previous Work} \label{Methcomparison}

Several studies have used statistical models to associate the onsite occurrence of induced seismicity with recorded SWDs, we make relevant comparisons below. Many of the models that appeared in the literature are modified versions of the classical Gutenberg–Richter law (GR law) which expresses the relationship between earthquake magnitude and earthquake frequency for natural earthquakes \cite{GRLaw}. \citeA{Langenbruch&Zoback2016HowWillSeismicityinOKRespondtoDecreasedSWD} modified the classical GR law to comprise contributions from cumulative injection volume, and seismogenic index \cite{shapiro2010seismogenic}, as expressed in (\ref{LangGR}) of \ref{app:previouswork}. While their primary goal was leveraging time-series data to predict future earthquakes from cumulative injection volume, one can extract an estimated association between injection activity and induced seismicity as a function of their model terms (\ref{LangGRAlter}). The quantity extracted from their analysis corresponds to an average increase in cumulative number of earthquakes with a unit increase in cumulative injection volume, which is a linear relationship that could be qualitatively compared with the type of relative rates estimated here. However, this association should be contrasted with our relative risk where the former is not data-driven in the sense it is obtained with outside information, while the latter is estimated with data at hand. 

\citeA{Grigoratos2020EarthquakesHindcasting} extended the approaches of \citeA{Langenbruch&Zoback2016HowWillSeismicityinOKRespondtoDecreasedSWD} to account for background tectonic activity rate.  An estimate of the linear association between injection activity and seismicity can be similarly extracted from expression (\ref{Grigoratos2020}) in \ref{app:previouswork}. They also provide a hypothesis testing procedure to test whether this linear association is significant across discretized grids over Oklahoma. Both \citeA{Grigoratos2020EarthquakesHindcasting} and \citeA{Langenbruch&Zoback2016HowWillSeismicityinOKRespondtoDecreasedSWD} rely heavily on the specification of the GR law and do not explicitly consider the possibility of (time-varying) confounding. Neither provides a means to assess statistical uncertainty of the estimated relationship between injection volume and seismicity.

\citeA{mcclure2017identifying} offered an analysis, using time-series data across California and Oklahoma, to model Poisson rates of seismicity related to injection activity, without adhering to the GR law. They focused on estimating the strength of relationship between injection in a given year on earthquakes in the same year over discretized grids. In particular, their focus on a contemporaneous effect precludes the need to carefully account for time-varying confounding, although they alluded to this possibility at a sub-annual time scale. They tested the null hypothesis that injection volume is not related to seismicity, however, their model does not entail any clear way to extract an interpretable quantity describing the strenght of association between injection volume and earthquakes.

These studies are an embodiment of the recent literature that statistically investigates the relationship between induced seismicity and SWD injection. Different model formulations present different considerations, however, none of these works provide interpretable estimates of the relationship between injection and seismicity with uncertainty, nor do they explicitly address or account for the possibility of time-varying confounding. Despite qualitative agreement of the estimated relationships between injection volume and seismicity in ours and previous work, our numerical estimates are not directly comparable because of (a) significantly different seismicity rates observed in Oklahoma and California compared to DFW, and (b) our focus on modeling relative rates vs. linear associations. However, we highlight the type of bias that we are interested to adjust is agnostic of the statistical model specification, and relates more fundamentaly to the causal identification conditions described in Section \ref{IgnorabilityIdentificationTimeFixed} and \ref{identificationtimevaryingconfounding} that establish the circumnstances under which estimation of causal effects is possible. Furthermore, the present paper highlights that accounting for the evolving dynamics between time-varying treatments, intermediate state of the process and the final outcome of the process would furnish the available statistical tool-box in deal with time-varying confounding and broaden the viewpoints of many geoscience problems that evolve time which is ubiquitous.

\subsection{Future Work}

Causal inference methodology has become popular and led to important contributions in a variety of other disciplines including education, psychology, economics, epidemiology, medicine, sociology \cite{Covidfriedrich2020causal, Glass2013CausalInferenceinPublicHealth, imbens_rubin_2015, gil2018intelligent}. It has been mostly unexplored in geoscience \cite{perez2018causalgeoscience, massmann2021causalearthscience}. Observational geosciences studies involve complicated dependence structures and confounding that are difficult to fully account for. For instance, simplification of the complexity of the problem (e.g. assuming the SUTVA) can potentially invalidate the causal portion of the analysis \cite{carone&Dominici2020TheRoleofCausallyDrivenDataScience}.
For example, it is reasonable to suppose that some SWDs all contribute to the occurrence of earthquakes in nearby locations, therefore, it is implausible to isolate the cause of earthquakes to specific SWDs. Although we applied agglomerative clustering on SWDs to better satisfy the no interference assumption, this ad-hoc approach is less than ideal to bypass the need of novel statistical methodology innovation to accounting for interference. These are subject to future work with a recently developed bipartite interference network that specifically targets causal inference with interference with application in observational geosciences problems \cite{zigler2021bipartiteinterference, zigler2020bipartite, giffin2020generalizedPSwithSpatialInterference, Marrett2018CorrelationSpace}.

\pagebreak

\appendix

\section{} \label{secondDAG}

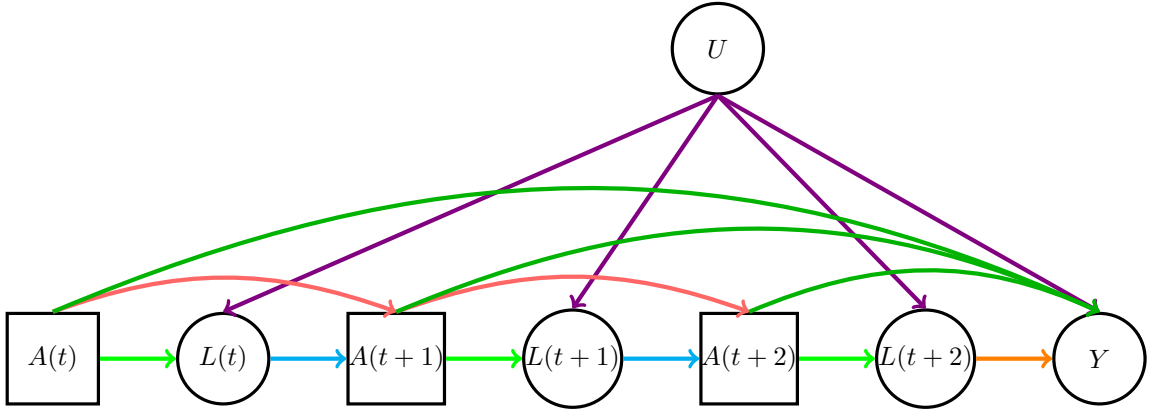
\begin{figure}[h] 
    \centering 
\begin{tikzpicture}[
roundnode/.style={circle, draw=black!, very thick, minimum size=12mm, inner sep=0pt},
squarednode/.style={rectangle, draw=black!, very thick, minimum size=12mm, inner sep=0pt},
]
\node[squarednode]      (Zt)          {$A(t)$};
\node[roundnode]        (Yt)         [right=of Zt]   {$L(t)$};
\node[squarednode]      (Ztplus1)       [right=of Yt]   {$A(t+1)$};
\node[roundnode]        (Ytplus1)       [right=of Ztplus1] {$L(t+1)$};
\node[squarednode]       (Ztplus2)       [right=of Ytplus1] {$A(t+2)$};
\node[roundnode]        (Ytplus2)       [right=of Ztplus2] {$L(t+2)$};
\node[roundnode]       (Y)       [right=of Ytplus2] {$Y$};
\node[roundnode]       (U)        [above right=of Ytplus1,yshift=2.2cm]{$U$}; 
\draw[ultra thick,green,->] (Zt.east) -- (Yt.west);
\draw[ultra thick,cyan,->] (Yt.east) -- (Ztplus1.west);
\draw[ultra thick,green,->] (Ztplus1.east) -- (Ytplus1.west);
\draw[ultra thick,cyan,->] (Ytplus1.east) -- (Ztplus2.west);
\draw[ultra thick,green,->] (Ztplus2.east) -- (Ytplus2.west);
\draw[ultra thick, orange, ->] (Ytplus2.east) --(Y.west);

\draw[ultra thick,violet,->] (U.south) -- (Yt.north);
\draw[ultra thick,violet,->] (U.south) -- (Ytplus1.north);
\draw[ultra thick,violet,->] (U.south) -- (Ytplus2.north);
\draw[ultra thick,violet,->] (U.south) -- (Y.north);
\draw[ultra thick,red!60,->] (Zt.north) to [out=20,in=160] (Ztplus1.north);
\draw[ultra thick,red!60,->] (Ztplus1.north) to [out=20,in=160] (Ztplus2.north);

\draw[ultra thick,black!30!green,->] (Zt.north) to [out=23,in=155] (Y.north);
\draw[ultra thick,black!30!green,->] (Ztplus1.north) to [out=23,in=155] (Y.north);
\draw[ultra thick,black!30!green,->] (Ztplus2.north) to [out=23,in=155] (Y.north);

\end{tikzpicture}
\caption{The DAG shows a complicated case of time-varying confounding with application in induced seismicity. The $A(t)$ denotes the monthly injection volume at time $t$, $L(t)$ denotes an intermediate indication of seismic propensity at time $t$, $U$ denotes time-fixed latent seismic risk that confounds $L(t)$ and $Y$, and $Y$ is the number of cumulative earthquakes at time $K+1$. The arrows represent established causal pathways in induced seismicity.}
\label{long dag 2}
\end{figure}

\section{}\label{datagenerationalgorithm}

\begin{algorithm}[H] 
\caption{Longitudinal data generation with time-varying confounding}
$\mbox{causal effect} = 0.001$\;
$\mbox{confounding} = 0.1$\;
$U \in (1,2,...,10)$\;
$M \gets 2000$\;
$N \gets 50$\;
$K \gets 8$\;
\mathleft \noindent
\begin{align} \label{bernP}
    \mbox{logit}[Pr(L_i(t) = 1| A_i(t), U_i)] = 0.14 * U_i + 1.1* \mathbbm{1}[A_i(t)>1000]
\end{align}

\mathleft \noindent
\begin{align}
    A(0) \sim & \mbox{Norm}(1000, 60) \\
    L(0)\sim & \mbox{Bern}(Pr(L(1) = 1| A(0), U)) 
\end{align}

\For{{$t \gets 1$ to $K$} }{

\noindent
\begin{align} \label{futureinjectionV1}
     A(t) \sim & \mbox{Norm}(A(t-1) - 55 * L(t-1) + 15, 60)
\end{align}

\noindent
\begin{align}
    L(t) \sim & \mbox{Bern}(Pr(L(t) = 1| (A(t), U) ))
\end{align} 
      
}

\noindent
\begin{align}
    cum(\Bar{A})& =  \sum^K_{t=1} A(t)\\ 
    cum(\Bar{L})& =  \sum^K_{t=1} L(t) \\
    Y &=  \mbox{Possion}(\mbox{exp}(\mbox{causal effect} * cum(\Bar{A}) + \mbox{confounding} * U))  \label{Ysimulated}
\end{align}

\textbf{return} $Y$, $cum(\Bar{A})$ and $cum(\Bar{L})$ for $i \in$  (1,50) as $\textbf{Y}$, $cum(\Bar{\textbf{A}})$ and $cum(\Bar{\textbf{L}})$

 \label{algorithm1}
\end{algorithm} 

\pagebreak

\section{} \label{simulationsensitivity}

\begin{figure}[H] 
    \centering
    \includegraphics[width=13cm]{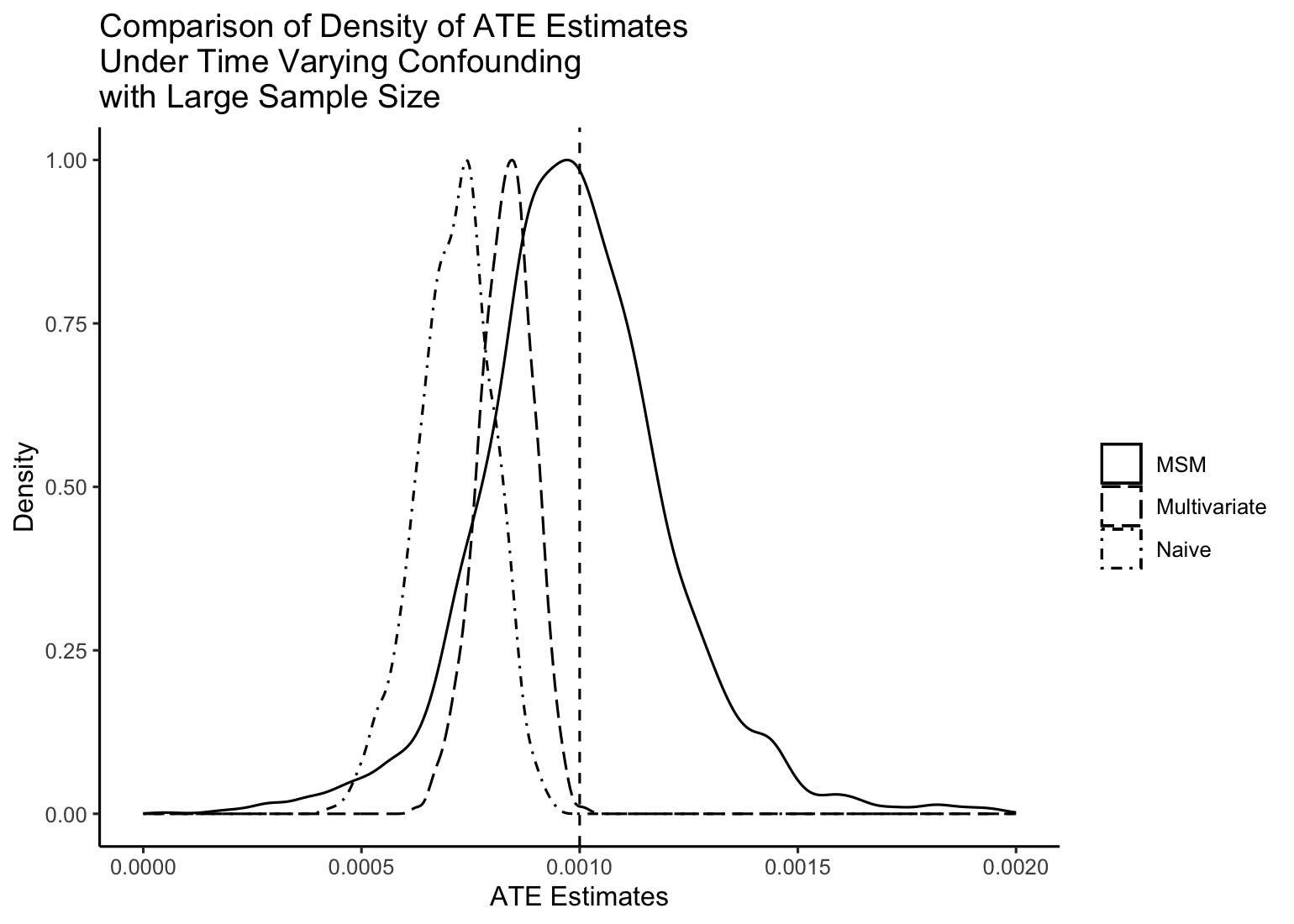}
    \caption{The figure shows the comparison of ATE estimates between three regression models across 2000 generated datasets with 500 clusters. The solid line is the density for the regression model with \textbf{ipw}, the dot dash line is the density for the naive regression model, and the long dash line is the density for the multivariate regression model. The dashed vertical line displays the true causal effect parameter used in the data generation.}
    \label{fig:Simulationestimatecomparison2}
\end{figure}

We repeat the simulation study outlined in Section \ref{simulation} with a larger sample size of 600 clusters. It is shown in Figure \ref{fig:Simulationestimatecomparison2} that the standard deviation of the estimated coefficients distribution shrinks as sample size increases. This is confirmed by comparing Table \ref{simulationtabl} with Table \ref{simulationtable2}. Having a larger sample size induces a smaller standard error which translates into tighter bounds for the confidence intervals for all three Poisson regression models. 

\begin{table}[h!]\label{simulationtable2}
\centering
\begin{tabular}{m{5cm} m{2cm} m{2cm} m{2cm}} 
\hline
Models & Average Point Estimates & Average Standard Error & Coverage\\
\hline
   Naive Regression Model  & 7.1e-04 & 8.81e-05 & 0.11 \\
   Multiple Regression Model  & 8.3e-04 &  6.24e-05  & 0.14 \\
   MSM with \textbf{ipw}  & 9.9e-04  & 0.000219 & 0.92 \\
\hline
\end{tabular} 
\caption{The average point estimate of regression coefficients, in thousands bbl, across 2000 datasets is shown, along with the average of standard errors for each Poisson regression, respectively. Furthermore, the biased estimation in point estimate has resulted in low coverage rate for naive and multiple Poisson regressions.}
\end{table}

\pagebreak

\section{} \label{app:previouswork}

The GR law has constants $a$ and $b$, the $M$ is the earthquake magnitude of completeness and the $N$ is the number of earthquakes: 
\begin{equation} \label{GRlaw}
    log_{10}[N_{\geq M}] = a - bM
\end{equation} 

\begin{equation} \label{LangGR}
    log_{10}[N_{\geq M}(t)] = a(t) - bM = log_{10}[V_{I}(t)] + \Sigma - bM
\end{equation}
This can be easily recognized after re-expressing their equation as: 
\begin{equation} \label{LangGRAlter} 
    N_{\geq M}(t) = V_{I}(t) * 10^{\Sigma - bM} 
\end{equation} 

where the quantity $10^{\Sigma - bM}$ essentially quantifies how the cumulative number of earthquakes respond to changes in cumulative injection volume over some specified time duration. Using their estimates of seismogenic indices equal to $-0.47$ and $-0.63$ and the $b$ equals to $1.41$ and $1.33$ for Central Oklahoma and West Oklahoma, respectively, one an recover. These amounts to $10^{-0.47 - 1.41*3}$ = 1.995e-5 and $10^{-0.63 - 1.33*3}$ = 2.399e-5 for Central Oklahoma and West Oklahoma, respectively. Therefore, \citeA{Langenbruch&Zoback2016HowWillSeismicityinOKRespondtoDecreasedSWD} estimated the average increase in cumulative number of earthquakes to be 19.95 and 23.99 for every additional 1 million cubic meter increase in cumulative injection volume, respectively.

Their re-formulation can be expressed as:
\begin{equation} \label{Grigoratos2020}
    N_{\geq M}(t) = 10^{a_{tec} - bM} + V_{I}(t) * 10^{\Sigma - bM}
\end{equation} where the $10^{a_{tec} - bM}$ can be viewed as an intercept in linear regression. 

\pagebreak
\section{} \label{realdatasensitivity}
We conduct a sensitivity analysis on number of clusters with $N=50$ and repeat the analysis with results shown in Table \ref{resultstable2}. Increasing the number of clusters can increase the statistical power of the analysis, however, the limited available earthquake catalog prevents uncovering any meaningful effects. We conclude the available earthquake catalog does not support for a fine grain scaled analysis. 

\begin{table}[H]
\centering
\begin{tabular}{m{2cm} m{2cm} m{2cm} m{2cm} m{2cm} m{2cm}}
\hline
Models & Parameter & Estimated Relative Risk in 1 MMbbl &  Standard Error & Wald Test Statistic &  P-values \\
\hline
   Naive    & $\gamma_1$ & 1.00 & 1.02e-02 & 0.20 & 0.84  \\
   Multivariate  & $\eta_1$ & 1.00 & 1.67e-02 & 0.28 & 0.78   \\
   MSM   & $\beta_1$ & 1.00 & 2.53e-02 & 0.24  &  0.81 \\
 \hline
 \end{tabular}
 \caption{The results table shows the estimated relative risk in 1 MMbbl, $exp(\beta_1 * 1e06)$, instead of the estimated regression coefficients, $\beta_1$, for the ease of interpretation. }
 \label{resultstable2}
 \end{table}

In particular, the time of data acquisition plays an important role on incorporating them in longitudinal analysis with time-varying confounding. For example, since many faults are mapped after the earthquakes had occurred and the mapping of faults is continuously embellished in time, it is inappropriate to denote the mapping of faults as a fixed pre-treatment confounder \cite{savvaidis2019texnet, Hennings2019Injection-inducedTexas, gao2019hydrologicalmodelinginDFW, DeShon2019NorthTexasEarthquakeSummary}.

\acknowledgments
We thank Louis Quinones and Heather DeShon for providing their databases of the Dallas Fort-Worth region. We thank Iason Grigoratos for helping to decluster the earthquake catalog. The financial support of the TexNet Research and the Center for Integrated Seismicity Research is gratefully acknowledged.
The earthquake catalog used in this paper is available from \citeA{DeShon2019NorthTexasEarthquakeSummary} and the saltwater disposal data is available from Texas Data Repository (\url{https://dataverse.tdl.org/dataset.xhtml?persistentId=doi:10.18738/T8/CEQEDF}). The gas well coordinates of the Fort-Worth region is provided by Jean-Philippe and Rebecca Gao from the Center for Integrated Seismicity Research of the Bureau of Economic Geology.


%
%

\bibliography{Bibliography}   

\begin{thebibliography}{}

\bibitem [\protect \citeauthoryear {%
Aldrich%
}{%
Aldrich%
}{%
{\protect \APACyear {1995}}%
}]{%
aldrich1995correlationdoesnotimplycausation}
\APACinsertmetastar {%
aldrich1995correlationdoesnotimplycausation}%
\begin{APACrefauthors}%
Aldrich, J.%
\end{APACrefauthors}%
\unskip\
\newblock
\APACrefYearMonthDay{1995}{11}{}.
\newblock
{\BBOQ}\APACrefatitle {Correlations Genuine and Spurious in Pearson and Yule} {Correlations genuine and spurious in pearson and yule}.{\BBCQ}
\newblock
\APACjournalVolNumPages{Statist. Sci.}{10}{4}{364--376}.
\newblock
\begin{APACrefURL} \url{https://doi.org/10.1214/ss/1177009870} \end{APACrefURL}
\newblock
\begin{APACrefDOI} \doi{10.1214/ss/1177009870} \end{APACrefDOI}
\PrintBackRefs{\CurrentBib}

\bibitem [\protect \citeauthoryear {%
Angrist%
, Imbens%
\BCBL {}\ \BBA {} Rubin%
}{%
Angrist%
\ \protect \BOthers {.}}{%
{\protect \APACyear {1996}}%
}]{%
ImbensandRubin1996}
\APACinsertmetastar {%
ImbensandRubin1996}%
\begin{APACrefauthors}%
Angrist, J\BPBI D.%
, Imbens, G\BPBI W.%
\BCBL {}\ \BBA {} Rubin, D\BPBI B.%
\end{APACrefauthors}%
\unskip\
\newblock
\APACrefYearMonthDay{1996}{}{}.
\newblock
{\BBOQ}\APACrefatitle {Identification of Causal Effects Using Instrumental Variables} {Identification of causal effects using instrumental variables}.{\BBCQ}
\newblock
\APACjournalVolNumPages{Journal of the American Statistical Association}{91}{434}{444--455}.
\newblock
\begin{APACrefURL} [{2023-03-02}]\url{http://www.jstor.org/stable/2291629} \end{APACrefURL}
\PrintBackRefs{\CurrentBib}

\bibitem [\protect \citeauthoryear {%
Carone%
, Dominici%
\BCBL {}\ \BBA {} Sheppard%
}{%
Carone%
\ \protect \BOthers {.}}{%
{\protect \APACyear {2020}}%
}]{%
carone&Dominici2020TheRoleofCausallyDrivenDataScience}
\APACinsertmetastar {%
carone&Dominici2020TheRoleofCausallyDrivenDataScience}%
\begin{APACrefauthors}%
Carone, M.%
, Dominici, F.%
\BCBL {}\ \BBA {} Sheppard, L.%
\end{APACrefauthors}%
\unskip\
\newblock
\APACrefYearMonthDay{2020}{}{}.
\newblock
{\BBOQ}\APACrefatitle {In pursuit of evidence in air pollution epidemiology: the role of causally driven data science} {In pursuit of evidence in air pollution epidemiology: the role of causally driven data science}.{\BBCQ}
\newblock
\APACjournalVolNumPages{Epidemiology}{31}{1}{1--6}.
\PrintBackRefs{\CurrentBib}

\bibitem [\protect \citeauthoryear {%
Cole%
\ \protect \BOthers {.}}{%
Cole%
\ \protect \BOthers {.}}{%
{\protect \APACyear {2010}}%
}]{%
cole2010coliderillustrating}
\APACinsertmetastar {%
cole2010coliderillustrating}%
\begin{APACrefauthors}%
Cole, S\BPBI R.%
, Platt, R\BPBI W.%
, Schisterman, E\BPBI F.%
, Chu, H.%
, Westreich, D.%
, Richardson, D.%
\BCBL {}\ \BBA {} Poole, C.%
\end{APACrefauthors}%
\unskip\
\newblock
\APACrefYearMonthDay{2010}{}{}.
\newblock
{\BBOQ}\APACrefatitle {Illustrating bias due to conditioning on a collider} {Illustrating bias due to conditioning on a collider}.{\BBCQ}
\newblock
\APACjournalVolNumPages{International journal of epidemiology}{39}{2}{417--420}.
\PrintBackRefs{\CurrentBib}

\bibitem [\protect \citeauthoryear {%
Cribari-Neto%
\ \BBA {} da Silva%
}{%
Cribari-Neto%
\ \BBA {} da Silva%
}{%
{\protect \APACyear {2011}}%
}]{%
cribari2011vcovHC4}
\APACinsertmetastar {%
cribari2011vcovHC4}%
\begin{APACrefauthors}%
Cribari-Neto, F.%
\BCBT {}\ \BBA {} da Silva, W\BPBI B.%
\end{APACrefauthors}%
\unskip\
\newblock
\APACrefYearMonthDay{2011}{}{}.
\newblock
{\BBOQ}\APACrefatitle {A new heteroskedasticity-consistent covariance matrix estimator for the linear regression model} {A new heteroskedasticity-consistent covariance matrix estimator for the linear regression model}.{\BBCQ}
\newblock
\APACjournalVolNumPages{AStA Advances in Statistical Analysis}{95}{}{129--146}.
\PrintBackRefs{\CurrentBib}

\bibitem [\protect \citeauthoryear {%
DeShon%
\ \protect \BOthers {.}}{%
DeShon%
\ \protect \BOthers {.}}{%
{\protect \APACyear {2019}}%
}]{%
DeShon2019NorthTexasEarthquakeSummary}
\APACinsertmetastar {%
DeShon2019NorthTexasEarthquakeSummary}%
\begin{APACrefauthors}%
DeShon, H\BPBI R.%
, Hayward, C\BPBI T.%
, Ogwari, P\BPBI O.%
, Quinones, L.%
, Sufri, O.%
, Stump, B.%
\BCBL {}\ \BBA {} Beatrice~Magnani, M.%
\end{APACrefauthors}%
\unskip\
\newblock
\APACrefYearMonthDay{2019}{}{}.
\newblock
{\BBOQ}\APACrefatitle {{Summary of the North Texas Earthquake Study Seismic Networks, 2013–2018}} {{Summary of the North Texas Earthquake Study Seismic Networks, 2013–2018}}.{\BBCQ}
\newblock
\APACjournalVolNumPages{Seismological Research Letters}{90}{1}{387--394}.
\newblock
\begin{APACrefDOI} \doi{10.1785/0220180269} \end{APACrefDOI}
\PrintBackRefs{\CurrentBib}

\bibitem [\protect \citeauthoryear {%
Dominici%
\ \BBA {} Zigler%
}{%
Dominici%
\ \BBA {} Zigler%
}{%
{\protect \APACyear {2017}}%
}]{%
dominici&Zigler2017GaugingEvidenceofCausalityinAirPollution}
\APACinsertmetastar {%
dominici&Zigler2017GaugingEvidenceofCausalityinAirPollution}%
\begin{APACrefauthors}%
Dominici, F.%
\BCBT {}\ \BBA {} Zigler, C.%
\end{APACrefauthors}%
\unskip\
\newblock
\APACrefYearMonthDay{2017}{}{}.
\newblock
{\BBOQ}\APACrefatitle {Best practices for gauging evidence of causality in air pollution epidemiology} {Best practices for gauging evidence of causality in air pollution epidemiology}.{\BBCQ}
\newblock
\APACjournalVolNumPages{American journal of epidemiology}{186}{12}{1303--1309}.
\PrintBackRefs{\CurrentBib}

\bibitem [\protect \citeauthoryear {%
Ellsworth%
}{%
Ellsworth%
}{%
{\protect \APACyear {2013}}%
}]{%
Ellsworth2013InjectionInducedEarthquakes}
\APACinsertmetastar {%
Ellsworth2013InjectionInducedEarthquakes}%
\begin{APACrefauthors}%
Ellsworth, W\BPBI L.%
\end{APACrefauthors}%
\unskip\
\newblock
\APACrefYearMonthDay{2013}{}{}.
\newblock
{\BBOQ}\APACrefatitle {Injection-Induced Earthquakes} {Injection-induced earthquakes}.{\BBCQ}
\newblock
\APACjournalVolNumPages{Science}{341}{6142}{}.
\newblock
\begin{APACrefURL} \url{https://science.sciencemag.org/content/341/6142/1225942} \end{APACrefURL}
\newblock
\begin{APACrefDOI} \doi{10.1126/science.1225942} \end{APACrefDOI}
\PrintBackRefs{\CurrentBib}

\bibitem [\protect \citeauthoryear {%
Elwert%
\ \BBA {} Winship%
}{%
Elwert%
\ \BBA {} Winship%
}{%
{\protect \APACyear {2014}}%
}]{%
elwert2014colliderendogenous}
\APACinsertmetastar {%
elwert2014colliderendogenous}%
\begin{APACrefauthors}%
Elwert, F.%
\BCBT {}\ \BBA {} Winship, C.%
\end{APACrefauthors}%
\unskip\
\newblock
\APACrefYearMonthDay{2014}{}{}.
\newblock
{\BBOQ}\APACrefatitle {Endogenous selection bias: The problem of conditioning on a collider variable} {Endogenous selection bias: The problem of conditioning on a collider variable}.{\BBCQ}
\newblock
\APACjournalVolNumPages{Annual review of sociology}{40}{}{31--53}.
\PrintBackRefs{\CurrentBib}

\bibitem [\protect \citeauthoryear {%
Fan%
, Eichhubl%
\BCBL {}\ \BBA {} Newell%
}{%
Fan%
\ \protect \BOthers {.}}{%
{\protect \APACyear {2019}}%
}]{%
Fan2019BasementEffects}
\APACinsertmetastar {%
Fan2019BasementEffects}%
\begin{APACrefauthors}%
Fan, Z.%
, Eichhubl, P.%
\BCBL {}\ \BBA {} Newell, P.%
\end{APACrefauthors}%
\unskip\
\newblock
\APACrefYearMonthDay{2019}{}{}.
\newblock
{\BBOQ}\APACrefatitle {{Basement Fault Reactivation by Fluid Injection Into Sedimentary Reservoirs: Poroelastic Effects}} {{Basement Fault Reactivation by Fluid Injection Into Sedimentary Reservoirs: Poroelastic Effects}}.{\BBCQ}
\newblock
\APACjournalVolNumPages{Journal of Geophysical Research: Solid Earth}{124}{7}{7354--7369}.
\newblock
\begin{APACrefDOI} \doi{10.1029/2018JB017062} \end{APACrefDOI}
\PrintBackRefs{\CurrentBib}

\bibitem [\protect \citeauthoryear {%
Fasola%
\ \protect \BOthers {.}}{%
Fasola%
\ \protect \BOthers {.}}{%
{\protect \APACyear {2019}}%
}]{%
Fasola2019HFstrategyinfluencetheprobabilityofEarthquakes}
\APACinsertmetastar {%
Fasola2019HFstrategyinfluencetheprobabilityofEarthquakes}%
\begin{APACrefauthors}%
Fasola, S\BPBI L.%
, Brudzinski, M\BPBI R.%
, Skoumal, R\BPBI J.%
, Langenkamp, T.%
, Currie, B\BPBI S.%
\BCBL {}\ \BBA {} Smart, K\BPBI J.%
\end{APACrefauthors}%
\unskip\
\newblock
\APACrefYearMonthDay{2019}{}{}.
\newblock
{\BBOQ}\APACrefatitle {Hydraulic Fracture Injection Strategy Influences the Probability of Earthquakes in the Eagle Ford Shale Play of South Texas} {Hydraulic fracture injection strategy influences the probability of earthquakes in the eagle ford shale play of south texas}.{\BBCQ}
\newblock
\APACjournalVolNumPages{Geophysical Research Letters}{46}{22}{12958-12967}.
\newblock
\begin{APACrefURL} \url{https://agupubs.onlinelibrary.wiley.com/doi/abs/10.1029/2019GL085167} \end{APACrefURL}
\newblock
\begin{APACrefDOI} \doi{https://doi.org/10.1029/2019GL085167} \end{APACrefDOI}
\PrintBackRefs{\CurrentBib}

\bibitem [\protect \citeauthoryear {%
Fitzmaurice%
, Davidian%
, Verbeke%
\BCBL {}\ \BBA {} Molenberghs%
}{%
Fitzmaurice%
\ \protect \BOthers {.}}{%
{\protect \APACyear {2008}}%
}]{%
longitudinaldataanalysis}
\APACinsertmetastar {%
longitudinaldataanalysis}%
\begin{APACrefauthors}%
Fitzmaurice, G.%
, Davidian, M.%
, Verbeke, G.%
\BCBL {}\ \BBA {} Molenberghs, G.%
\end{APACrefauthors}%
\unskip\
\newblock
\APACrefYear{2008}.
\newblock
\APACrefbtitle {Longitudinal data analysis} {Longitudinal data analysis}.
\newblock
\APACaddressPublisher{}{CRC press}.
\PrintBackRefs{\CurrentBib}

\bibitem [\protect \citeauthoryear {%
Freedman%
}{%
Freedman%
}{%
{\protect \APACyear {2006}}%
}]{%
freedman2006sandwich}
\APACinsertmetastar {%
freedman2006sandwich}%
\begin{APACrefauthors}%
Freedman, D\BPBI A.%
\end{APACrefauthors}%
\unskip\
\newblock
\APACrefYearMonthDay{2006}{}{}.
\newblock
{\BBOQ}\APACrefatitle {On the so-called “Huber sandwich estimator” and “robust standard errors”} {On the so-called “huber sandwich estimator” and “robust standard errors”}.{\BBCQ}
\newblock
\APACjournalVolNumPages{The American Statistician}{60}{4}{299--302}.
\PrintBackRefs{\CurrentBib}

\bibitem [\protect \citeauthoryear {%
Friedrich%
\ \BBA {} Friede%
}{%
Friedrich%
\ \BBA {} Friede%
}{%
{\protect \APACyear {2020}}%
}]{%
Covidfriedrich2020causal}
\APACinsertmetastar {%
Covidfriedrich2020causal}%
\begin{APACrefauthors}%
Friedrich, S.%
\BCBT {}\ \BBA {} Friede, T.%
\end{APACrefauthors}%
\unskip\
\newblock
\APACrefYearMonthDay{2020}{}{}.
\newblock
{\BBOQ}\APACrefatitle {Causal inference methods for small non-randomized studies: Methods and an application in COVID-19} {Causal inference methods for small non-randomized studies: Methods and an application in covid-19}.{\BBCQ}
\newblock
\APACjournalVolNumPages{Contemporary clinical trials}{99}{}{106213}.
\PrintBackRefs{\CurrentBib}

\bibitem [\protect \citeauthoryear {%
Frohlich%
\ \protect \BOthers {.}}{%
Frohlich%
\ \protect \BOthers {.}}{%
{\protect \APACyear {2016}}%
}]{%
Frohlich2016ATexas}
\APACinsertmetastar {%
Frohlich2016ATexas}%
\begin{APACrefauthors}%
Frohlich, C.%
, De~Shon, H.%
, Stump, B.%
, Hayward, C.%
, Hornbach, M.%
\BCBL {}\ \BBA {} Walter, J\BPBI I.%
\end{APACrefauthors}%
\unskip\
\newblock
\APACrefYearMonthDay{2016}{}{}.
\newblock
{\BBOQ}\APACrefatitle {{A historical review of induced Earthquakes in Texas}} {{A historical review of induced Earthquakes in Texas}}.{\BBCQ}
\newblock
\APACjournalVolNumPages{Seismological Research Letters}{87}{4}{1022--1038}.
\newblock
\begin{APACrefDOI} \doi{10.1785/0220160016} \end{APACrefDOI}
\PrintBackRefs{\CurrentBib}

\bibitem [\protect \citeauthoryear {%
Frohlich%
\ \protect \BOthers {.}}{%
Frohlich%
\ \protect \BOthers {.}}{%
{\protect \APACyear {2020}}%
}]{%
Frohlich2020OnsetArray}
\APACinsertmetastar {%
Frohlich2020OnsetArray}%
\begin{APACrefauthors}%
Frohlich, C.%
, Hayward, C.%
, Rosenblit, J.%
, Aiken, C.%
, Hennings, P.%
, Savvaidis, A.%
\BDBL {}DeShon, H\BPBI R.%
\end{APACrefauthors}%
\unskip\
\newblock
\APACrefYearMonthDay{2020}{}{}.
\newblock
{\BBOQ}\APACrefatitle {{Onset and Cause of Increased Seismic Activity Near Pecos, West Texas, United States, From Observations at the Lajitas TXAR Seismic Array}} {{Onset and Cause of Increased Seismic Activity Near Pecos, West Texas, United States, From Observations at the Lajitas TXAR Seismic Array}}.{\BBCQ}
\newblock
\APACjournalVolNumPages{Journal of Geophysical Research: Solid Earth}{125}{1}{1--14}.
\newblock
\begin{APACrefURL} \url{https://doi.org/10.1029/2019JB017737} \end{APACrefURL}
\newblock
\begin{APACrefDOI} \doi{10.1029/2019JB017737} \end{APACrefDOI}
\PrintBackRefs{\CurrentBib}

\bibitem [\protect \citeauthoryear {%
Gao%
, Pelletier%
, Horne%
, Nicot%
\BCBL {}\ \BBA {} Hennings%
}{%
Gao%
\ \protect \BOthers {.}}{%
{\protect \APACyear {2019}}%
}]{%
gao2019hydrologicalmodelinginDFW}
\APACinsertmetastar {%
gao2019hydrologicalmodelinginDFW}%
\begin{APACrefauthors}%
Gao, R.%
, Pelletier, I.%
, Horne, E.%
, Nicot, J\BHBI P.%
\BCBL {}\ \BBA {} Hennings, P.%
\end{APACrefauthors}%
\unskip\
\newblock
\APACrefYearMonthDay{2019}{}{}.
\newblock
{\BBOQ}\APACrefatitle {Basin-scale hydrogeological modeling of the Fort Worth basin Ellenburger group for pore pressure characterization} {Basin-scale hydrogeological modeling of the fort worth basin ellenburger group for pore pressure characterization}.{\BBCQ}
\newblock
\BIn{} \APACrefbtitle {AGU Fall Meeting Abstracts} {Agu fall meeting abstracts}\ (\BVOL\ 2019, \BPGS\ H53C--01).
\PrintBackRefs{\CurrentBib}

\bibitem [\protect \citeauthoryear {%
Giffin%
, Reich%
, Yang%
\BCBL {}\ \BBA {} Rappold%
}{%
Giffin%
\ \protect \BOthers {.}}{%
{\protect \APACyear {2020}}%
}]{%
giffin2020generalizedPSwithSpatialInterference}
\APACinsertmetastar {%
giffin2020generalizedPSwithSpatialInterference}%
\begin{APACrefauthors}%
Giffin, A.%
, Reich, B.%
, Yang, S.%
\BCBL {}\ \BBA {} Rappold, A.%
\end{APACrefauthors}%
\unskip\
\newblock
\APACrefYearMonthDay{2020}{}{}.
\newblock
\APACrefbtitle {Generalized propensity score approach to causal inference with spatial interference.} {Generalized propensity score approach to causal inference with spatial interference.}
\PrintBackRefs{\CurrentBib}

\bibitem [\protect \citeauthoryear {%
Gil%
\ \protect \BOthers {.}}{%
Gil%
\ \protect \BOthers {.}}{%
{\protect \APACyear {2018}}%
}]{%
gil2018intelligent}
\APACinsertmetastar {%
gil2018intelligent}%
\begin{APACrefauthors}%
Gil, Y.%
, Pierce, S\BPBI A.%
, Babaie, H.%
, Banerjee, A.%
, Borne, K.%
, Bust, G.%
\BDBL {}others%
\end{APACrefauthors}%
\unskip\
\newblock
\APACrefYearMonthDay{2018}{}{}.
\newblock
{\BBOQ}\APACrefatitle {Intelligent systems for geosciences: an essential research agenda} {Intelligent systems for geosciences: an essential research agenda}.{\BBCQ}
\newblock
\APACjournalVolNumPages{Communications of the ACM}{62}{1}{76--84}.
\PrintBackRefs{\CurrentBib}

\bibitem [\protect \citeauthoryear {%
Glass%
, Goodman%
, Hernán%
\BCBL {}\ \BBA {} Samet%
}{%
Glass%
\ \protect \BOthers {.}}{%
{\protect \APACyear {2013}}%
}]{%
Glass2013CausalInferenceinPublicHealth}
\APACinsertmetastar {%
Glass2013CausalInferenceinPublicHealth}%
\begin{APACrefauthors}%
Glass, T\BPBI A.%
, Goodman, S\BPBI N.%
, Hernán, M\BPBI A.%
\BCBL {}\ \BBA {} Samet, J\BPBI M.%
\end{APACrefauthors}%
\unskip\
\newblock
\APACrefYearMonthDay{2013}{}{}.
\newblock
{\BBOQ}\APACrefatitle {Causal Inference in Public Health} {Causal inference in public health}.{\BBCQ}
\newblock
\APACjournalVolNumPages{Annual Review of Public Health}{34}{1}{61-75}.
\newblock
\begin{APACrefURL} \url{https://doi.org/10.1146/annurev-publhealth-031811-124606} \end{APACrefURL}
\newblock
\APACrefnote{PMID: 23297653}
\newblock
\begin{APACrefDOI} \doi{10.1146/annurev-publhealth-031811-124606} \end{APACrefDOI}
\PrintBackRefs{\CurrentBib}

\bibitem [\protect \citeauthoryear {%
Grigoratos%
, Rathje%
, Bazzurro%
\BCBL {}\ \BBA {} Savvaidis%
}{%
Grigoratos%
\ \protect \BOthers {.}}{%
{\protect \APACyear {2020}}%
{\protect \APACexlab {{\protect \BCnt {1}}}}}]{%
Grigoratos2020EarthquakesForecasting}
\APACinsertmetastar {%
Grigoratos2020EarthquakesForecasting}%
\begin{APACrefauthors}%
Grigoratos, I.%
, Rathje, E.%
, Bazzurro, P.%
\BCBL {}\ \BBA {} Savvaidis, A.%
\end{APACrefauthors}%
\unskip\
\newblock
\APACrefYearMonthDay{2020{\protect \BCnt {1}}}{}{}.
\newblock
{\BBOQ}\APACrefatitle {{Earthquakes Induced by Wastewater Injection, Part II: Statistical Evaluation of Causal Factors and Seismicity Rate Forecasting}} {{Earthquakes Induced by Wastewater Injection, Part II: Statistical Evaluation of Causal Factors and Seismicity Rate Forecasting}}.{\BBCQ}
\newblock
\APACjournalVolNumPages{Bulletin of the Seismological Society of America}{110}{5}{2483--2497}.
\newblock
\begin{APACrefDOI} \doi{10.1785/0120200079} \end{APACrefDOI}
\PrintBackRefs{\CurrentBib}

\bibitem [\protect \citeauthoryear {%
Grigoratos%
, Rathje%
, Bazzurro%
\BCBL {}\ \BBA {} Savvaidis%
}{%
Grigoratos%
\ \protect \BOthers {.}}{%
{\protect \APACyear {2020}}%
{\protect \APACexlab {{\protect \BCnt {2}}}}}]{%
Grigoratos2020EarthquakesHindcasting}
\APACinsertmetastar {%
Grigoratos2020EarthquakesHindcasting}%
\begin{APACrefauthors}%
Grigoratos, I.%
, Rathje, E.%
, Bazzurro, P.%
\BCBL {}\ \BBA {} Savvaidis, A.%
\end{APACrefauthors}%
\unskip\
\newblock
\APACrefYearMonthDay{2020{\protect \BCnt {2}}}{}{}.
\newblock
{\BBOQ}\APACrefatitle {{Earthquakes Induced by Wastewater Injection, Part I: Model Development and Hindcasting}} {{Earthquakes Induced by Wastewater Injection, Part I: Model Development and Hindcasting}}.{\BBCQ}
\newblock
\APACjournalVolNumPages{Bulletin of the Seismological Society of America}{110}{5}{2466--2482}.
\newblock
\begin{APACrefDOI} \doi{10.1785/0120200078} \end{APACrefDOI}
\PrintBackRefs{\CurrentBib}

\bibitem [\protect \citeauthoryear {%
Gutenberg%
\ \BBA {} Richter%
}{%
Gutenberg%
\ \BBA {} Richter%
}{%
{\protect \APACyear {1956}}%
}]{%
GRLaw}
\APACinsertmetastar {%
GRLaw}%
\begin{APACrefauthors}%
Gutenberg, B.%
\BCBT {}\ \BBA {} Richter, C\BPBI F.%
\end{APACrefauthors}%
\unskip\
\newblock
\APACrefYearMonthDay{1956}{}{}.
\newblock
{\BBOQ}\APACrefatitle {Earthquake magnitude, intensity, energy, and acceleration: (Second paper)} {Earthquake magnitude, intensity, energy, and acceleration: (second paper)}.{\BBCQ}
\newblock
\APACjournalVolNumPages{Bulletin of the seismological society of America}{46}{2}{105--145}.
\PrintBackRefs{\CurrentBib}

\bibitem [\protect \citeauthoryear {%
Hennings%
\ \protect \BOthers {.}}{%
Hennings%
\ \protect \BOthers {.}}{%
{\protect \APACyear {2019}}%
}]{%
Hennings2019Injection-inducedTexas}
\APACinsertmetastar {%
Hennings2019Injection-inducedTexas}%
\begin{APACrefauthors}%
Hennings, P\BPBI H.%
, Snee, J\BPBI E\BPBI L.%
, Osmond, J\BPBI L.%
, Deshon, H\BPBI R.%
, Dommisse, R.%
, Horne, E.%
\BDBL {}Zoback, M\BPBI D.%
\end{APACrefauthors}%
\unskip\
\newblock
\APACrefYearMonthDay{2019}{}{}.
\newblock
{\BBOQ}\APACrefatitle {{Injection-Induced Seismicity and Fault-Slip Potential in the Fort Worth Basin, Texas}} {{Injection-Induced Seismicity and Fault-Slip Potential in the Fort Worth Basin, Texas}}.{\BBCQ}
\newblock
\APACjournalVolNumPages{Bulletin of the Seismological Society of America}{109}{5}{1615--1634}.
\newblock
\begin{APACrefDOI} \doi{10.1785/0120190017} \end{APACrefDOI}
\PrintBackRefs{\CurrentBib}

\bibitem [\protect \citeauthoryear {%
Hirano%
\ \BBA {} Imbens%
}{%
Hirano%
\ \BBA {} Imbens%
}{%
{\protect \APACyear {2004}}%
}]{%
hirano2004generalizedpropensity}
\APACinsertmetastar {%
hirano2004generalizedpropensity}%
\begin{APACrefauthors}%
Hirano, K.%
\BCBT {}\ \BBA {} Imbens, G\BPBI W.%
\end{APACrefauthors}%
\unskip\
\newblock
\APACrefYearMonthDay{2004}{}{}.
\newblock
{\BBOQ}\APACrefatitle {The propensity score with continuous treatments} {The propensity score with continuous treatments}.{\BBCQ}
\newblock
\APACjournalVolNumPages{Applied Bayesian modeling and causal inference from incomplete-data perspectives}{226164}{}{73--84}.
\PrintBackRefs{\CurrentBib}

\bibitem [\protect \citeauthoryear {%
Holland%
}{%
Holland%
}{%
{\protect \APACyear {1986}}%
}]{%
Holland1986statisticsandcausalinference}
\APACinsertmetastar {%
Holland1986statisticsandcausalinference}%
\begin{APACrefauthors}%
Holland, P\BPBI W.%
\end{APACrefauthors}%
\unskip\
\newblock
\APACrefYearMonthDay{1986}{}{}.
\newblock
{\BBOQ}\APACrefatitle {Statistics and Causal Inference} {Statistics and causal inference}.{\BBCQ}
\newblock
\APACjournalVolNumPages{Journal of the American Statistical Association}{81}{396}{945--960}.
\newblock
\begin{APACrefURL} \url{http://www.jstor.org/stable/2289064} \end{APACrefURL}
\PrintBackRefs{\CurrentBib}

\bibitem [\protect \citeauthoryear {%
Hornbach%
\ \protect \BOthers {.}}{%
Hornbach%
\ \protect \BOthers {.}}{%
{\protect \APACyear {2015}}%
}]{%
Hornbach2015CausalTexas}
\APACinsertmetastar {%
Hornbach2015CausalTexas}%
\begin{APACrefauthors}%
Hornbach, M\BPBI J.%
, Deshon, H\BPBI R.%
, Ellsworth, W\BPBI L.%
, Stump, B\BPBI W.%
, Hayward, C.%
, Frohlich, C.%
\BDBL {}Luetgert, J\BPBI H.%
\end{APACrefauthors}%
\unskip\
\newblock
\APACrefYearMonthDay{2015}{}{}.
\newblock
{\BBOQ}\APACrefatitle {{Causal Factors for Seismicity near Azle, Texas}} {{Causal Factors for Seismicity near Azle, Texas}}.{\BBCQ}
\newblock
\APACjournalVolNumPages{Nature Communications}{6}{}{1--11}.
\newblock
\begin{APACrefURL} \url{http://dx.doi.org/10.1038/ncomms7728} \end{APACrefURL}
\newblock
\begin{APACrefDOI} \doi{10.1038/ncomms7728} \end{APACrefDOI}
\PrintBackRefs{\CurrentBib}

\bibitem [\protect \citeauthoryear {%
Hornbach%
\ \protect \BOthers {.}}{%
Hornbach%
\ \protect \BOthers {.}}{%
{\protect \APACyear {2016}}%
}]{%
Hornbach2016EllenburgerTexas}
\APACinsertmetastar {%
Hornbach2016EllenburgerTexas}%
\begin{APACrefauthors}%
Hornbach, M\BPBI J.%
, Jones, M.%
, Scales, M.%
, DeShon, H\BPBI R.%
, Magnani, M\BPBI B.%
, Frohlich, C.%
\BDBL {}Layton, M.%
\end{APACrefauthors}%
\unskip\
\newblock
\APACrefYearMonthDay{2016}{}{}.
\newblock
{\BBOQ}\APACrefatitle {{Ellenburger wastewater injection and seismicity in North Texas}} {{Ellenburger wastewater injection and seismicity in North Texas}}.{\BBCQ}
\newblock
\APACjournalVolNumPages{Physics of the Earth and Planetary Interiors}{261}{}{54--68}.
\newblock
\begin{APACrefURL} \url{http://dx.doi.org/10.1016/j.pepi.2016.06.012} \end{APACrefURL}
\newblock
\begin{APACrefDOI} \doi{10.1016/j.pepi.2016.06.012} \end{APACrefDOI}
\PrintBackRefs{\CurrentBib}

\bibitem [\protect \citeauthoryear {%
Imbens%
\ \BBA {} Rubin%
}{%
Imbens%
\ \BBA {} Rubin%
}{%
{\protect \APACyear {2015}}%
}]{%
imbens_rubin_2015}
\APACinsertmetastar {%
imbens_rubin_2015}%
\begin{APACrefauthors}%
Imbens, G\BPBI W.%
\BCBT {}\ \BBA {} Rubin, D\BPBI B.%
\end{APACrefauthors}%
\unskip\
\newblock
\APACrefYear{2015}.
\newblock
\APACrefbtitle {Causal Inference for Statistics, Social, and Biomedical Sciences: An Introduction} {Causal inference for statistics, social, and biomedical sciences: An introduction}.
\newblock
\APACaddressPublisher{}{Cambridge University Press}.
\newblock
\begin{APACrefDOI} \doi{10.1017/CBO9781139025751} \end{APACrefDOI}
\PrintBackRefs{\CurrentBib}

\bibitem [\protect \citeauthoryear {%
Justinic%
, Stump%
, Hayward%
\BCBL {}\ \BBA {} Frohlich%
}{%
Justinic%
\ \protect \BOthers {.}}{%
{\protect \APACyear {2013}}%
}]{%
Justinic2013Analysis2010}
\APACinsertmetastar {%
Justinic2013Analysis2010}%
\begin{APACrefauthors}%
Justinic, A\BPBI H.%
, Stump, B.%
, Hayward, C.%
\BCBL {}\ \BBA {} Frohlich, C.%
\end{APACrefauthors}%
\unskip\
\newblock
\APACrefYearMonthDay{2013}{}{}.
\newblock
{\BBOQ}\APACrefatitle {{Analysis of the Cleburne, Texas, Earthquake Sequence from June 2009 to June 2010}} {{Analysis of the Cleburne, Texas, Earthquake Sequence from June 2009 to June 2010}}.{\BBCQ}
\newblock
\APACjournalVolNumPages{Bulletin of the Seismological Society of America}{103}{6}{3083--3093}.
\newblock
\begin{APACrefDOI} \doi{10.1785/0120120336} \end{APACrefDOI}
\PrintBackRefs{\CurrentBib}

\bibitem [\protect \citeauthoryear {%
Keranen%
, Savage%
, Abers%
\BCBL {}\ \BBA {} Cochran%
}{%
Keranen%
\ \protect \BOthers {.}}{%
{\protect \APACyear {2013}}%
}]{%
keranen2013potentiallyInducedEarthquakesinOK}
\APACinsertmetastar {%
keranen2013potentiallyInducedEarthquakesinOK}%
\begin{APACrefauthors}%
Keranen, K\BPBI M.%
, Savage, H\BPBI M.%
, Abers, G\BPBI A.%
\BCBL {}\ \BBA {} Cochran, E\BPBI S.%
\end{APACrefauthors}%
\unskip\
\newblock
\APACrefYearMonthDay{2013}{}{}.
\newblock
{\BBOQ}\APACrefatitle {Potentially induced earthquakes in Oklahoma, USA: Links between wastewater injection and the 2011 Mw 5.7 earthquake sequence} {Potentially induced earthquakes in oklahoma, usa: Links between wastewater injection and the 2011 mw 5.7 earthquake sequence}.{\BBCQ}
\newblock
\APACjournalVolNumPages{Geology}{41}{6}{699--702}.
\PrintBackRefs{\CurrentBib}

\bibitem [\protect \citeauthoryear {%
Keranen%
, Weingarten%
, Abers%
, Bekins%
\BCBL {}\ \BBA {} Ge%
}{%
Keranen%
\ \protect \BOthers {.}}{%
{\protect \APACyear {2014}}%
}]{%
Keranen2014SharpIncreaseinOKSeismicityRatesBYSWD}
\APACinsertmetastar {%
Keranen2014SharpIncreaseinOKSeismicityRatesBYSWD}%
\begin{APACrefauthors}%
Keranen, K\BPBI M.%
, Weingarten, M.%
, Abers, G\BPBI A.%
, Bekins, B\BPBI A.%
\BCBL {}\ \BBA {} Ge, S.%
\end{APACrefauthors}%
\unskip\
\newblock
\APACrefYearMonthDay{2014}{}{}.
\newblock
{\BBOQ}\APACrefatitle {Sharp increase in central Oklahoma seismicity since 2008 induced by massive wastewater injection} {Sharp increase in central oklahoma seismicity since 2008 induced by massive wastewater injection}.{\BBCQ}
\newblock
\APACjournalVolNumPages{Science}{345}{6195}{448--451}.
\newblock
\begin{APACrefURL} \url{https://science.sciencemag.org/content/345/6195/448} \end{APACrefURL}
\newblock
\begin{APACrefDOI} \doi{10.1126/science.1255802} \end{APACrefDOI}
\PrintBackRefs{\CurrentBib}

\bibitem [\protect \citeauthoryear {%
Langenbruch%
\ \BBA {} Zoback%
}{%
Langenbruch%
\ \BBA {} Zoback%
}{%
{\protect \APACyear {2016}}%
}]{%
Langenbruch&Zoback2016HowWillSeismicityinOKRespondtoDecreasedSWD}
\APACinsertmetastar {%
Langenbruch&Zoback2016HowWillSeismicityinOKRespondtoDecreasedSWD}%
\begin{APACrefauthors}%
Langenbruch, C.%
\BCBT {}\ \BBA {} Zoback, M\BPBI D.%
\end{APACrefauthors}%
\unskip\
\newblock
\APACrefYearMonthDay{2016}{}{}.
\newblock
{\BBOQ}\APACrefatitle {How will induced seismicity in Oklahoma respond to decreased saltwater injection rates?} {How will induced seismicity in oklahoma respond to decreased saltwater injection rates?}{\BBCQ}
\newblock
\APACjournalVolNumPages{Science Advances}{2}{11}{}.
\newblock
\begin{APACrefDOI} \doi{10.1126/sciadv.1601542} \end{APACrefDOI}
\PrintBackRefs{\CurrentBib}

\bibitem [\protect \citeauthoryear {%
Langenbruch%
\ \BBA {} Zoback%
}{%
Langenbruch%
\ \BBA {} Zoback%
}{%
{\protect \APACyear {2017}}%
}]{%
Langenbruch2017ResponseRates}
\APACinsertmetastar {%
Langenbruch2017ResponseRates}%
\begin{APACrefauthors}%
Langenbruch, C.%
\BCBT {}\ \BBA {} Zoback, M\BPBI D.%
\end{APACrefauthors}%
\unskip\
\newblock
\APACrefYearMonthDay{2017}{}{}.
\newblock
{\BBOQ}\APACrefatitle {{Response to Comment on “How will induced seismicity in Oklahoma respond to decreased saltwater injection rates?”}} {{Response to Comment on “How will induced seismicity in Oklahoma respond to decreased saltwater injection rates?”}}.{\BBCQ}
\newblock
\APACjournalVolNumPages{Science Advances}{3}{8}{1--10}.
\newblock
\begin{APACrefDOI} \doi{10.1126/sciadv.aao2277} \end{APACrefDOI}
\PrintBackRefs{\CurrentBib}

\bibitem [\protect \citeauthoryear {%
Lund~Snee%
\ \BBA {} Zoback%
}{%
Lund~Snee%
\ \BBA {} Zoback%
}{%
{\protect \APACyear {2016}}%
}]{%
Snee&Zoback2016StateofStressinTX}
\APACinsertmetastar {%
Snee&Zoback2016StateofStressinTX}%
\begin{APACrefauthors}%
Lund~Snee, J\BHBI E.%
\BCBT {}\ \BBA {} Zoback, M\BPBI D.%
\end{APACrefauthors}%
\unskip\
\newblock
\APACrefYearMonthDay{2016}{}{}.
\newblock
{\BBOQ}\APACrefatitle {State of stress in Texas: Implications for induced seismicity} {State of stress in texas: Implications for induced seismicity}.{\BBCQ}
\newblock
\APACjournalVolNumPages{Geophysical Research Letters}{43}{19}{10,208-10,214}.
\newblock
\begin{APACrefURL} \url{https://agupubs.onlinelibrary.wiley.com/doi/abs/10.1002/2016GL070974} \end{APACrefURL}
\newblock
\begin{APACrefDOI} \doi{https://doi.org/10.1002/2016GL070974} \end{APACrefDOI}
\PrintBackRefs{\CurrentBib}

\bibitem [\protect \citeauthoryear {%
Marrett%
, Gale%
, G{\'{o}}mez%
\BCBL {}\ \BBA {} Laubach%
}{%
Marrett%
\ \protect \BOthers {.}}{%
{\protect \APACyear {2018}}%
}]{%
Marrett2018CorrelationSpace}
\APACinsertmetastar {%
Marrett2018CorrelationSpace}%
\begin{APACrefauthors}%
Marrett, R.%
, Gale, J\BPBI F.%
, G{\'{o}}mez, L\BPBI A.%
\BCBL {}\ \BBA {} Laubach, S\BPBI E.%
\end{APACrefauthors}%
\unskip\
\newblock
\APACrefYearMonthDay{2018}{}{}.
\newblock
{\BBOQ}\APACrefatitle {{Correlation Analysis of Fracture Arrangement in Space}} {{Correlation Analysis of Fracture Arrangement in Space}}.{\BBCQ}
\newblock
\APACjournalVolNumPages{Journal of Structural Geology}{108}{}{16--33}.
\newblock
\begin{APACrefDOI} \doi{10.1016/j.jsg.2017.06.012} \end{APACrefDOI}
\PrintBackRefs{\CurrentBib}

\bibitem [\protect \citeauthoryear {%
Massmann%
, Gentine%
\BCBL {}\ \BBA {} Runge%
}{%
Massmann%
\ \protect \BOthers {.}}{%
{\protect \APACyear {2021}}%
}]{%
massmann2021causalearthscience}
\APACinsertmetastar {%
massmann2021causalearthscience}%
\begin{APACrefauthors}%
Massmann, A.%
, Gentine, P.%
\BCBL {}\ \BBA {} Runge, J.%
\end{APACrefauthors}%
\unskip\
\newblock
\APACrefYearMonthDay{2021}{}{}.
\newblock
{\BBOQ}\APACrefatitle {Causal inference for process understanding in Earth sciences} {Causal inference for process understanding in earth sciences}.{\BBCQ}
\newblock
\APACjournalVolNumPages{arXiv preprint arXiv:2105.00912}{}{}{}.
\PrintBackRefs{\CurrentBib}

\bibitem [\protect \citeauthoryear {%
McClure%
, Gibson%
, Chiu%
\BCBL {}\ \BBA {} Ranganath%
}{%
McClure%
\ \protect \BOthers {.}}{%
{\protect \APACyear {2017}}%
}]{%
mcclure2017identifying}
\APACinsertmetastar {%
mcclure2017identifying}%
\begin{APACrefauthors}%
McClure, M.%
, Gibson, R.%
, Chiu, K\BHBI K.%
\BCBL {}\ \BBA {} Ranganath, R.%
\end{APACrefauthors}%
\unskip\
\newblock
\APACrefYearMonthDay{2017}{}{}.
\newblock
{\BBOQ}\APACrefatitle {Identifying potentially induced seismicity and assessing statistical significance in Oklahoma and California} {Identifying potentially induced seismicity and assessing statistical significance in oklahoma and california}.{\BBCQ}
\newblock
\APACjournalVolNumPages{Journal of Geophysical Research: Solid Earth}{122}{3}{2153--2172}.
\PrintBackRefs{\CurrentBib}

\bibitem [\protect \citeauthoryear {%
Ogwari%
, DeShon%
\BCBL {}\ \BBA {} Hornbach%
}{%
Ogwari%
\ \protect \BOthers {.}}{%
{\protect \APACyear {2018}}%
}]{%
Ogwari&Deshon2018DFWEarthquakeSequence}
\APACinsertmetastar {%
Ogwari&Deshon2018DFWEarthquakeSequence}%
\begin{APACrefauthors}%
Ogwari, P\BPBI O.%
, DeShon, H\BPBI R.%
\BCBL {}\ \BBA {} Hornbach, M\BPBI J.%
\end{APACrefauthors}%
\unskip\
\newblock
\APACrefYearMonthDay{2018}{}{}.
\newblock
{\BBOQ}\APACrefatitle {The Dallas-Fort Worth Airport Earthquake Sequence: Seismicity Beyond Injection Period} {The dallas-fort worth airport earthquake sequence: Seismicity beyond injection period}.{\BBCQ}
\newblock
\APACjournalVolNumPages{Journal of Geophysical Research: Solid Earth}{123}{1}{553-563}.
\newblock
\begin{APACrefURL} \url{https://agupubs.onlinelibrary.wiley.com/doi/abs/10.1002/2017JB015003} \end{APACrefURL}
\newblock
\begin{APACrefDOI} \doi{https://doi.org/10.1002/2017JB015003} \end{APACrefDOI}
\PrintBackRefs{\CurrentBib}

\bibitem [\protect \citeauthoryear {%
Pearl%
}{%
Pearl%
}{%
{\protect \APACyear {2019}}%
}]{%
pearl2019seventoolsofcausalinference}
\APACinsertmetastar {%
pearl2019seventoolsofcausalinference}%
\begin{APACrefauthors}%
Pearl, J.%
\end{APACrefauthors}%
\unskip\
\newblock
\APACrefYearMonthDay{2019}{}{}.
\newblock
{\BBOQ}\APACrefatitle {The seven tools of causal inference, with reflections on machine learning} {The seven tools of causal inference, with reflections on machine learning}.{\BBCQ}
\newblock
\APACjournalVolNumPages{Communications of the ACM}{62}{3}{54--60}.
\PrintBackRefs{\CurrentBib}

\bibitem [\protect \citeauthoryear {%
P{\'e}rez-Suay%
\ \BBA {} Camps-Valls%
}{%
P{\'e}rez-Suay%
\ \BBA {} Camps-Valls%
}{%
{\protect \APACyear {2018}}%
}]{%
perez2018causalgeoscience}
\APACinsertmetastar {%
perez2018causalgeoscience}%
\begin{APACrefauthors}%
P{\'e}rez-Suay, A.%
\BCBT {}\ \BBA {} Camps-Valls, G.%
\end{APACrefauthors}%
\unskip\
\newblock
\APACrefYearMonthDay{2018}{}{}.
\newblock
{\BBOQ}\APACrefatitle {Causal inference in geoscience and remote sensing from observational data} {Causal inference in geoscience and remote sensing from observational data}.{\BBCQ}
\newblock
\APACjournalVolNumPages{IEEE Transactions on Geoscience and Remote Sensing}{57}{3}{1502--1513}.
\PrintBackRefs{\CurrentBib}

\bibitem [\protect \citeauthoryear {%
L.~Quinones%
\ \protect \BOthers {.}}{%
L.~Quinones%
\ \protect \BOthers {.}}{%
{\protect \APACyear {2019}}%
}]{%
Quinones2019TrackingCatalog}
\APACinsertmetastar {%
Quinones2019TrackingCatalog}%
\begin{APACrefauthors}%
Quinones, L.%
, Deshon, H\BPBI R.%
, Jeong, S.%
, Ogwari, P.%
, Sufri, O.%
, Holt, M\BPBI M.%
\BCBL {}\ \BBA {} Kwong, K\BPBI B.%
\end{APACrefauthors}%
\unskip\
\newblock
\APACrefYearMonthDay{2019}{}{}.
\newblock
{\BBOQ}\APACrefatitle {{Tracking Induced Seismicity in the Fort Worth Basin: A Summary of the 2008–2018 North Texas Earthquake Study Catalog}} {{Tracking Induced Seismicity in the Fort Worth Basin: A Summary of the 2008–2018 North Texas Earthquake Study Catalog}}.{\BBCQ}
\newblock
\APACjournalVolNumPages{Bulletin of the Seismological Society of America}{109}{4}{1203--1216}.
\newblock
\begin{APACrefDOI} \doi{10.1785/0120190057} \end{APACrefDOI}
\PrintBackRefs{\CurrentBib}

\bibitem [\protect \citeauthoryear {%
L\BPBI A.~Quinones%
, DeShon%
, Magnani%
\BCBL {}\ \BBA {} Frohlich%
}{%
L\BPBI A.~Quinones%
\ \protect \BOthers {.}}{%
{\protect \APACyear {2018}}%
}]{%
Louis&Deshon2018StressOrientationsinDFW}
\APACinsertmetastar {%
Louis&Deshon2018StressOrientationsinDFW}%
\begin{APACrefauthors}%
Quinones, L\BPBI A.%
, DeShon, H\BPBI R.%
, Magnani, M\BPBI B.%
\BCBL {}\ \BBA {} Frohlich, C.%
\end{APACrefauthors}%
\unskip\
\newblock
\APACrefYearMonthDay{2018}{04}{}.
\newblock
{\BBOQ}\APACrefatitle {Stress Orientations in the Fort Worth Basin, Texas, Determined from Earthquake Focal Mechanisms} {Stress orientations in the fort worth basin, texas, determined from earthquake focal mechanisms}.{\BBCQ}
\newblock
\APACjournalVolNumPages{Bulletin of the Seismological Society of America}{108}{3A}{1124-1132}.
\newblock
\begin{APACrefURL} \url{https://doi.org/10.1785/0120170337} \end{APACrefURL}
\newblock
\begin{APACrefDOI} \doi{10.1785/0120170337} \end{APACrefDOI}
\PrintBackRefs{\CurrentBib}

\bibitem [\protect \citeauthoryear {%
Reich%
\ \protect \BOthers {.}}{%
Reich%
\ \protect \BOthers {.}}{%
{\protect \APACyear {2020}}%
}]{%
Reich2020Areviewofspatialcausalinferencemethodsforenvironmentalandepidemiologicalapplications}
\APACinsertmetastar {%
Reich2020Areviewofspatialcausalinferencemethodsforenvironmentalandepidemiologicalapplications}%
\begin{APACrefauthors}%
Reich, B.%
, Yang, S.%
, Guan, Y.%
, Giffin, A.%
, Miller, M\BPBI J.%
\BCBL {}\ \BBA {} Rappold, A.%
\end{APACrefauthors}%
\unskip\
\newblock
\APACrefYearMonthDay{2020}{}{}.
\newblock
{\BBOQ}\APACrefatitle {A review of spatial causal inference methods for environmental and epidemiological applications} {A review of spatial causal inference methods for environmental and epidemiological applications}.{\BBCQ}
\newblock
\APACjournalVolNumPages{arXiv: Methodology}{}{}{}.
\PrintBackRefs{\CurrentBib}

\bibitem [\protect \citeauthoryear {%
Robins%
}{%
Robins%
}{%
{\protect \APACyear {1997}}%
}]{%
robins1997marginal}
\APACinsertmetastar {%
robins1997marginal}%
\begin{APACrefauthors}%
Robins, J\BPBI M.%
\end{APACrefauthors}%
\unskip\
\newblock
\APACrefYearMonthDay{1997}{}{}.
\newblock
{\BBOQ}\APACrefatitle {Marginal structural models} {Marginal structural models}.{\BBCQ}
\newblock

\PrintBackRefs{\CurrentBib}

\bibitem [\protect \citeauthoryear {%
Robins%
, Hernan%
\BCBL {}\ \BBA {} Brumback%
}{%
Robins%
\ \protect \BOthers {.}}{%
{\protect \APACyear {2000}}%
}]{%
robins2000marginal}
\APACinsertmetastar {%
robins2000marginal}%
\begin{APACrefauthors}%
Robins, J\BPBI M.%
, Hernan, M\BPBI A.%
\BCBL {}\ \BBA {} Brumback, B.%
\end{APACrefauthors}%
\unskip\
\newblock
\APACrefYearMonthDay{2000}{}{}.
\newblock
{\BBOQ}\APACrefatitle {Marginal structural models and causal inference in epidemiology} {Marginal structural models and causal inference in epidemiology}.{\BBCQ}
\newblock
\APACjournalVolNumPages{Epidemiology}{}{}{550--560}.
\PrintBackRefs{\CurrentBib}

\bibitem [\protect \citeauthoryear {%
Savvaidis%
, Lomax%
\BCBL {}\ \BBA {} Breton%
}{%
Savvaidis%
\ \protect \BOthers {.}}{%
{\protect \APACyear {2020}}%
}]{%
savvaidis2020induced}
\APACinsertmetastar {%
savvaidis2020induced}%
\begin{APACrefauthors}%
Savvaidis, A.%
, Lomax, A.%
\BCBL {}\ \BBA {} Breton, C.%
\end{APACrefauthors}%
\unskip\
\newblock
\APACrefYearMonthDay{2020}{}{}.
\newblock
{\BBOQ}\APACrefatitle {Induced seismicity in the Delaware Basin, West Texas, is caused by hydraulic fracturing and wastewater disposal} {Induced seismicity in the delaware basin, west texas, is caused by hydraulic fracturing and wastewater disposal}.{\BBCQ}
\newblock
\APACjournalVolNumPages{Bulletin of the Seismological Society of America}{110}{5}{2225--2241}.
\PrintBackRefs{\CurrentBib}

\bibitem [\protect \citeauthoryear {%
Savvaidis%
, Young%
, Huang%
\BCBL {}\ \BBA {} Lomax%
}{%
Savvaidis%
\ \protect \BOthers {.}}{%
{\protect \APACyear {2019}}%
}]{%
savvaidis2019texnet}
\APACinsertmetastar {%
savvaidis2019texnet}%
\begin{APACrefauthors}%
Savvaidis, A.%
, Young, B.%
, Huang, G\BHBI c\BPBI D.%
\BCBL {}\ \BBA {} Lomax, A.%
\end{APACrefauthors}%
\unskip\
\newblock
\APACrefYearMonthDay{2019}{}{}.
\newblock
{\BBOQ}\APACrefatitle {TexNet: A statewide seismological network in Texas} {Texnet: A statewide seismological network in texas}.{\BBCQ}
\newblock
\APACjournalVolNumPages{Seismological Research Letters}{90}{4}{1702--1715}.
\PrintBackRefs{\CurrentBib}

\bibitem [\protect \citeauthoryear {%
Scales%
\ \protect \BOthers {.}}{%
Scales%
\ \protect \BOthers {.}}{%
{\protect \APACyear {2017}}%
}]{%
Scales2017ATexas}
\APACinsertmetastar {%
Scales2017ATexas}%
\begin{APACrefauthors}%
Scales, M\BPBI M.%
, DeShon, H\BPBI R.%
, Magnani, M\BPBI B.%
, Walter, J\BPBI I.%
, Quinones, L.%
, Pratt, T\BPBI L.%
\BCBL {}\ \BBA {} Hornbach, M\BPBI J.%
\end{APACrefauthors}%
\unskip\
\newblock
\APACrefYearMonthDay{2017}{}{}.
\newblock
{\BBOQ}\APACrefatitle {{A Decade of Induced Slip on the Causative Fault of the 2015 Mw 4.0 Venus Earthquake, Northeast Johnson County, Texas}} {{A Decade of Induced Slip on the Causative Fault of the 2015 Mw 4.0 Venus Earthquake, Northeast Johnson County, Texas}}.{\BBCQ}
\newblock
\APACjournalVolNumPages{Journal of Geophysical Research: Solid Earth}{122}{10}{7879--7894}.
\newblock
\begin{APACrefDOI} \doi{10.1002/2017JB014460} \end{APACrefDOI}
\PrintBackRefs{\CurrentBib}

\bibitem [\protect \citeauthoryear {%
Schoenball%
\ \BBA {} Ellsworth%
}{%
Schoenball%
\ \BBA {} Ellsworth%
}{%
{\protect \APACyear {2017}}%
}]{%
Schoenball&Ellsworth2017Assessmentofspatiotemporalevolutionoffaultactivation}
\APACinsertmetastar {%
Schoenball&Ellsworth2017Assessmentofspatiotemporalevolutionoffaultactivation}%
\begin{APACrefauthors}%
Schoenball, M.%
\BCBT {}\ \BBA {} Ellsworth, W\BPBI L.%
\end{APACrefauthors}%
\unskip\
\newblock
\APACrefYearMonthDay{2017}{}{}.
\newblock
{\BBOQ}\APACrefatitle {A Systematic Assessment of the Spatiotemporal Evolution of Fault Activation Through Induced Seismicity in Oklahoma and Southern Kansas} {A systematic assessment of the spatiotemporal evolution of fault activation through induced seismicity in oklahoma and southern kansas}.{\BBCQ}
\newblock
\APACjournalVolNumPages{Journal of Geophysical Research: Solid Earth}{122}{12}{10,189-10,206}.
\newblock
\begin{APACrefURL} \url{https://agupubs.onlinelibrary.wiley.com/doi/abs/10.1002/2017JB014850} \end{APACrefURL}
\newblock
\begin{APACrefDOI} \doi{https://doi.org/10.1002/2017JB014850} \end{APACrefDOI}
\PrintBackRefs{\CurrentBib}

\bibitem [\protect \citeauthoryear {%
Shapiro%
, Dinske%
, Langenbruch%
\BCBL {}\ \BBA {} Wenzel%
}{%
Shapiro%
\ \protect \BOthers {.}}{%
{\protect \APACyear {2010}}%
}]{%
shapiro2010seismogenic}
\APACinsertmetastar {%
shapiro2010seismogenic}%
\begin{APACrefauthors}%
Shapiro, S\BPBI A.%
, Dinske, C.%
, Langenbruch, C.%
\BCBL {}\ \BBA {} Wenzel, F.%
\end{APACrefauthors}%
\unskip\
\newblock
\APACrefYearMonthDay{2010}{}{}.
\newblock
{\BBOQ}\APACrefatitle {Seismogenic index and magnitude probability of earthquakes induced during reservoir fluid stimulations} {Seismogenic index and magnitude probability of earthquakes induced during reservoir fluid stimulations}.{\BBCQ}
\newblock
\APACjournalVolNumPages{The Leading Edge}{29}{3}{304--309}.
\PrintBackRefs{\CurrentBib}

\bibitem [\protect \citeauthoryear {%
Walsh%
\ \BBA {} Zoback%
}{%
Walsh%
\ \BBA {} Zoback%
}{%
{\protect \APACyear {2015}}%
}]{%
Walsh2015OklahomasDisposal}
\APACinsertmetastar {%
Walsh2015OklahomasDisposal}%
\begin{APACrefauthors}%
Walsh, F\BPBI R.%
\BCBT {}\ \BBA {} Zoback, M\BPBI D.%
\end{APACrefauthors}%
\unskip\
\newblock
\APACrefYearMonthDay{2015}{}{}.
\newblock
{\BBOQ}\APACrefatitle {{Oklahoma's recent earthquakes and saltwater disposal}} {{Oklahoma's recent earthquakes and saltwater disposal}}.{\BBCQ}
\newblock
\APACjournalVolNumPages{Science Advances}{1}{5}{1--10}.
\newblock
\begin{APACrefDOI} \doi{10.1126/sciadv.1500195} \end{APACrefDOI}
\PrintBackRefs{\CurrentBib}

\bibitem [\protect \citeauthoryear {%
Weingarten%
, Ge%
, Godt%
, Bekins%
\BCBL {}\ \BBA {} Rubinstein%
}{%
Weingarten%
\ \protect \BOthers {.}}{%
{\protect \APACyear {2015}}%
}]{%
Weingarten2015High-rateSeismicity}
\APACinsertmetastar {%
Weingarten2015High-rateSeismicity}%
\begin{APACrefauthors}%
Weingarten, M.%
, Ge, S.%
, Godt, J\BPBI W.%
, Bekins, B\BPBI A.%
\BCBL {}\ \BBA {} Rubinstein, J\BPBI L.%
\end{APACrefauthors}%
\unskip\
\newblock
\APACrefYearMonthDay{2015}{}{}.
\newblock
{\BBOQ}\APACrefatitle {{High-rate injection is associated with the increase in U.S. mid-continent seismicity}} {{High-rate injection is associated with the increase in U.S. mid-continent seismicity}}.{\BBCQ}
\newblock
\APACjournalVolNumPages{Science}{348}{6241}{1336--1340}.
\newblock
\begin{APACrefDOI} \doi{10.1126/science.aab1345} \end{APACrefDOI}
\PrintBackRefs{\CurrentBib}

\bibitem [\protect \citeauthoryear {%
White%
}{%
White%
}{%
{\protect \APACyear {1980}}%
}]{%
white1980heteroskedasticity}
\APACinsertmetastar {%
white1980heteroskedasticity}%
\begin{APACrefauthors}%
White, H.%
\end{APACrefauthors}%
\unskip\
\newblock
\APACrefYearMonthDay{1980}{}{}.
\newblock
{\BBOQ}\APACrefatitle {A heteroskedasticity-consistent covariance matrix estimator and a direct test for heteroskedasticity} {A heteroskedasticity-consistent covariance matrix estimator and a direct test for heteroskedasticity}.{\BBCQ}
\newblock
\APACjournalVolNumPages{Econometrica: journal of the Econometric Society}{}{}{817--838}.
\PrintBackRefs{\CurrentBib}

\bibitem [\protect \citeauthoryear {%
Wooldridge%
}{%
Wooldridge%
}{%
{\protect \APACyear {2007}}%
}]{%
wooldridge2007iptwVariance}
\APACinsertmetastar {%
wooldridge2007iptwVariance}%
\begin{APACrefauthors}%
Wooldridge, J\BPBI M.%
\end{APACrefauthors}%
\unskip\
\newblock
\APACrefYearMonthDay{2007}{}{}.
\newblock
{\BBOQ}\APACrefatitle {Inverse probability weighted estimation for general missing data problems} {Inverse probability weighted estimation for general missing data problems}.{\BBCQ}
\newblock
\APACjournalVolNumPages{Journal of econometrics}{141}{2}{1281--1301}.
\PrintBackRefs{\CurrentBib}

\bibitem [\protect \citeauthoryear {%
Young%
, Hern{\'a}n%
, Picciotto%
\BCBL {}\ \BBA {} Robins%
}{%
Young%
\ \protect \BOthers {.}}{%
{\protect \APACyear {2010}}%
}]{%
young2010relation}
\APACinsertmetastar {%
young2010relation}%
\begin{APACrefauthors}%
Young, J\BPBI G.%
, Hern{\'a}n, M\BPBI A.%
, Picciotto, S.%
\BCBL {}\ \BBA {} Robins, J\BPBI M.%
\end{APACrefauthors}%
\unskip\
\newblock
\APACrefYearMonthDay{2010}{}{}.
\newblock
{\BBOQ}\APACrefatitle {Relation between three classes of structural models for the effect of a time-varying exposure on survival} {Relation between three classes of structural models for the effect of a time-varying exposure on survival}.{\BBCQ}
\newblock
\APACjournalVolNumPages{Lifetime data analysis}{16}{}{71--84}.
\PrintBackRefs{\CurrentBib}

\bibitem [\protect \citeauthoryear {%
Zhai%
\ \BBA {} Shirzaei%
}{%
Zhai%
\ \BBA {} Shirzaei%
}{%
{\protect \APACyear {2018}}%
}]{%
Zhai2018FluidTexas}
\APACinsertmetastar {%
Zhai2018FluidTexas}%
\begin{APACrefauthors}%
Zhai, G.%
\BCBT {}\ \BBA {} Shirzaei, M.%
\end{APACrefauthors}%
\unskip\
\newblock
\APACrefYearMonthDay{2018}{}{}.
\newblock
{\BBOQ}\APACrefatitle {{Fluid Injection and Time-Dependent Seismic Hazard in the Barnett Shale, Texas}} {{Fluid Injection and Time-Dependent Seismic Hazard in the Barnett Shale, Texas}}.{\BBCQ}
\newblock
\APACjournalVolNumPages{Geophysical Research Letters}{45}{10}{4743--4753}.
\newblock
\begin{APACrefDOI} \doi{10.1029/2018GL077696} \end{APACrefDOI}
\PrintBackRefs{\CurrentBib}

\bibitem [\protect \citeauthoryear {%
Zhang%
, Hudgens%
\BCBL {}\ \BBA {} Halloran%
}{%
Zhang%
\ \protect \BOthers {.}}{%
{\protect \APACyear {2023}}%
}]{%
zhang2023propensity}
\APACinsertmetastar {%
zhang2023propensity}%
\begin{APACrefauthors}%
Zhang, B.%
, Hudgens, M\BPBI G.%
\BCBL {}\ \BBA {} Halloran, M\BPBI E.%
\end{APACrefauthors}%
\unskip\
\newblock
\APACrefYearMonthDay{2023}{}{}.
\newblock
{\BBOQ}\APACrefatitle {Propensity Score in the Face of Interference: Discussion of Rosenbaum and Rubin (1983)} {Propensity score in the face of interference: Discussion of rosenbaum and rubin (1983)}.{\BBCQ}
\newblock
\APACjournalVolNumPages{Observational Studies}{9}{1}{125--131}.
\PrintBackRefs{\CurrentBib}

\bibitem [\protect \citeauthoryear {%
C.~Zigler%
, Forastiere%
\BCBL {}\ \BBA {} Mealli%
}{%
C.~Zigler%
\ \protect \BOthers {.}}{%
{\protect \APACyear {2020}}%
}]{%
zigler2020bipartite}
\APACinsertmetastar {%
zigler2020bipartite}%
\begin{APACrefauthors}%
Zigler, C.%
, Forastiere, L.%
\BCBL {}\ \BBA {} Mealli, F.%
\end{APACrefauthors}%
\unskip\
\newblock
\APACrefYearMonthDay{2020}{}{}.
\newblock
\APACrefbtitle {Bipartite Interference and Air Pollution Transport: Estimating Health Effects of Power Plant Interventions.} {Bipartite interference and air pollution transport: Estimating health effects of power plant interventions.}
\PrintBackRefs{\CurrentBib}

\bibitem [\protect \citeauthoryear {%
C\BPBI M.~Zigler%
\ \BBA {} Papadogeorgou%
}{%
C\BPBI M.~Zigler%
\ \BBA {} Papadogeorgou%
}{%
{\protect \APACyear {2021}}%
}]{%
zigler2021bipartiteinterference}
\APACinsertmetastar {%
zigler2021bipartiteinterference}%
\begin{APACrefauthors}%
Zigler, C\BPBI M.%
\BCBT {}\ \BBA {} Papadogeorgou, G.%
\end{APACrefauthors}%
\unskip\
\newblock
\APACrefYearMonthDay{2021}{}{}.
\newblock
{\BBOQ}\APACrefatitle {{Bipartite Causal Inference with Interference}} {{Bipartite Causal Inference with Interference}}.{\BBCQ}
\newblock
\APACjournalVolNumPages{Statistical Science}{36}{1}{109 -- 123}.
\newblock
\begin{APACrefURL} \url{https://doi.org/10.1214/19-STS749} \end{APACrefURL}
\newblock
\begin{APACrefDOI} \doi{10.1214/19-STS749} \end{APACrefDOI}
\PrintBackRefs{\CurrentBib}

\end{thebibliography}

%
%
%
%
%

\end{document}